\documentclass[%
 aip,
 amsmath,amssymb,
 reprint,%
]{revtex4-2}

\usepackage{graphicx}
\usepackage{dcolumn}
\usepackage{bm}

\usepackage[utf8]{inputenc}
\usepackage[T1]{fontenc}
\usepackage{mathptmx}
\usepackage{etoolbox}
\usepackage{colortbl}
\usepackage{xcolor}

\makeatletter
\def\@email#1#2{%
 \endgroup
 \patchcmd{\titleblock@produce}
  {\frontmatter@RRAPformat}
  {\frontmatter@RRAPformat{\produce@RRAP{*#1\href{mailto:#2}{#2}}}\frontmatter@RRAPformat}
  {}{}
}%
\makeatother
\DeclareUnicodeCharacter{3B4}{$\delta$}
\DeclareUnicodeCharacter{2212}{-}

\begin{document}

\preprint{AIP/123-QED}
\newcommand{\CT}{Cr$_2$Te$_3$}
\newcommand{\CxT}{Cr$_{1+x}$Te$_2$}
\newcommand{\Tc}{$T_\mathrm{C}$}

\title{Magnetic evolution of Cr$_2$Te$_3$ epitaxially grown on graphene with post-growth annealing}

\author{Quentin Guillet}
\affiliation{Université Grenoble Alpes, CEA, CNRS, IRIG-SPINTEC, 38000 Grenoble, France
}
\author{Hervé Boukari}
\affiliation{Université Grenoble Alpes, CNRS, Institut N\'eel, 38000 Grenoble, France
}
\author{Fadi Choueikani}
\affiliation{Synchrotron SOLEIL, L'Orme des Merisiers, 91190 Saint-Aubin, France
}
\author{Philippe Ohresser}
\affiliation{Synchrotron SOLEIL, L'Orme des Merisiers, 91190 Saint-Aubin, France
}
\author{Abdelkarim Ouerghi}
\affiliation{Université Paris-Saclay, CNRS, Centre de Nanosciences et de Nanotechnologies, Palaiseau, France
}
\author{Florie Mesple}
\affiliation{Université Grenoble Alpes, CEA, CNRS, IRIG-PHELIQS, 38000 Grenoble, France
}
\author{Vincent T. Renard}
\affiliation{Université Grenoble Alpes, CEA, CNRS, IRIG-PHELIQS, 38000 Grenoble, France
}
\author{Jean-François Jacquot}
\affiliation{Université Grenoble Alpes, CEA, CNRS, IRIG-SYMMES, 38000 Grenoble, France
}
\author{Denis Jalabert}
\affiliation{Université Grenoble Alpes, CEA, IRIG-MEM, 38000 Grenoble, France
}
\author{Céline Vergnaud}
\affiliation{Université Grenoble Alpes, CEA, CNRS, IRIG-SPINTEC, 38000 Grenoble, France
}
\author{Frédéric Bonell}
\affiliation{Université Grenoble Alpes, CEA, CNRS, IRIG-SPINTEC, 38000 Grenoble, France
}
\author{Alain Marty}
\affiliation{Université Grenoble Alpes, CEA, CNRS, IRIG-SPINTEC, 38000 Grenoble, France
}
\author{Matthieu Jamet}
\affiliation{Université Grenoble Alpes, CEA, CNRS, IRIG-SPINTEC, 38000 Grenoble, France
}
\email{matthieu.jamet@cea.fr}
\date{\today}

\begin{abstract}
Two-dimensional and van der Waals ferromagnets are ideal platform to study low dimensional magnetism and proximity effects in van der Waals heterostructures. Their ultimate two dimensional character offers also the opportunity to easily adjust their magnetic properties using strain or electric fields. Among 2D ferromagnets, the \CxT\ compounds with $x$=0-1 are very promising because their magnetic properties depend on the amount of self-intercalated Cr atoms between pure CrTe$_2$ layers and the Curie temperature (\Tc) can reach room temperature for certain compositions. Here, we investigate the evolution of the composition, structural and magnetic properties of thin Cr$_{1.33}$Te$_2$ (\CT) films epitaxially grown on graphene upon annealing. We observe a transition above 450°C from the Cr$_{1.33}$Te$_2$ phase with perpendicular magnetic anisotropy and a \Tc\ of 180 K to a composition close to Cr$_{1.39}$Te$_2$ with in-plane magnetic anisotropy and a \Tc\ of 240-250 K. This phase remains stable up to 650°C above which a pure Cr film starts to form. This work demonstrates the complex interplay between intercalated Cr, lattice parameters and magnetic properties in \CxT\ compounds.
\end{abstract}

\maketitle
\begin{figure*}[ht]
    \begin{center}
    \includegraphics[width=17.2cm]{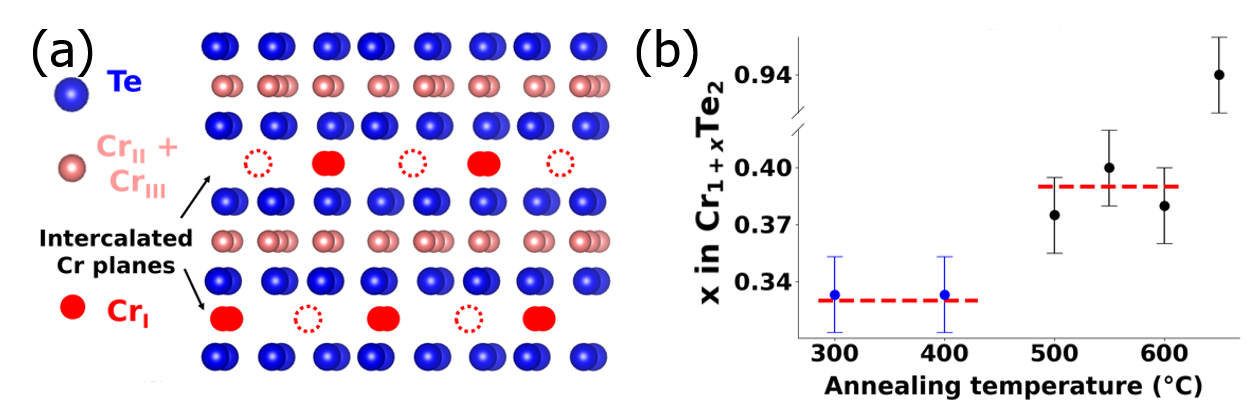}
    \caption{(a) Crystal structure of \CxT. Every second Cr layer is only partially filled, as evidenced by the dashed empty circles. The filling factor of these intercalated Cr planes corresponds to the $x$ number. (b) Evolution of the film stoichiometry as a function of the annealing temperature. Blue dots are for Cr$_{1.33}$Te$_2$ and black dots for other phases. Red dotted lines are visual guides indicating areas with little composition variation as a function of T$_A$.}
\label{Fig1}
\end{center}
\end{figure*}
\begin{figure*}[ht]
    \begin{center}
    \includegraphics[width=17.2cm]{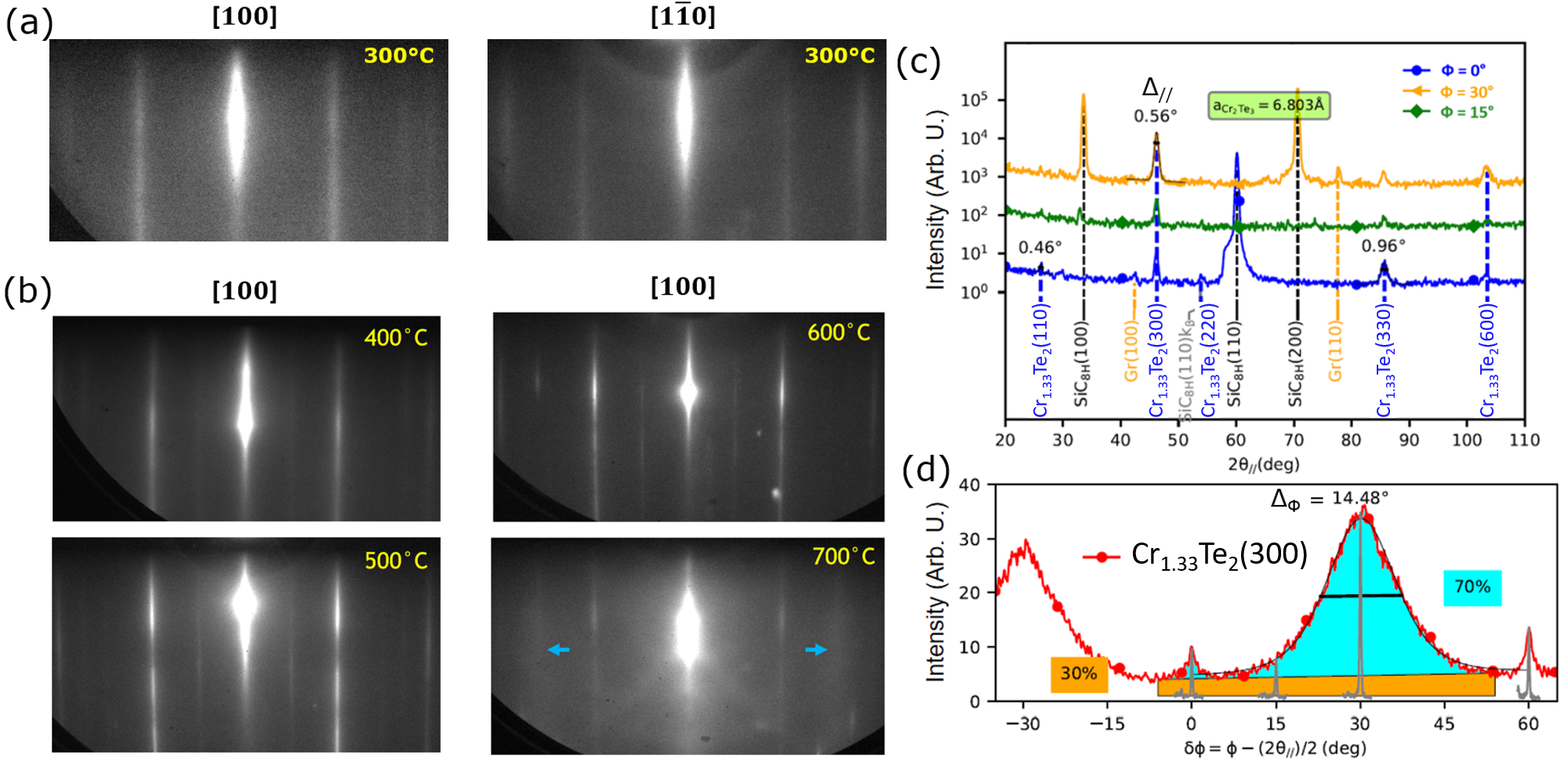}
    \caption{(a) RHEED images after deposition of 5 ML of Cr$_{1.33}$Te$_2$ on graphene/SiC at 300°C along the two high-symmetry crystalline directions rotated by 30° from each other. (b) RHEED images recorded along the [100] crystal direction after annealing of Cr$_{1.33}$Te$_2$ at the temperature indicated in the top right corner. Blue arrows on the last image indicate the position of graphene (100) diffraction rods. (c) Radial XRD spectra of 5 layers of Cr$_{1.33}$Te$_2$ annealed at 450°C. (d) Azimuthal scan of the same sample performed on the Cr$_{1.33}$Te$_2$ (300) Bragg peak. The isotropic (orange) and anisotropic (light blue) areas stand for the poly-crystalline and single crystalline contributions to the diffraction peaks.
    }
\label{Fig2}
\end{center}
\end{figure*}
\begin{figure*}[ht]
    \begin{center}
    \includegraphics[width=17.2cm]{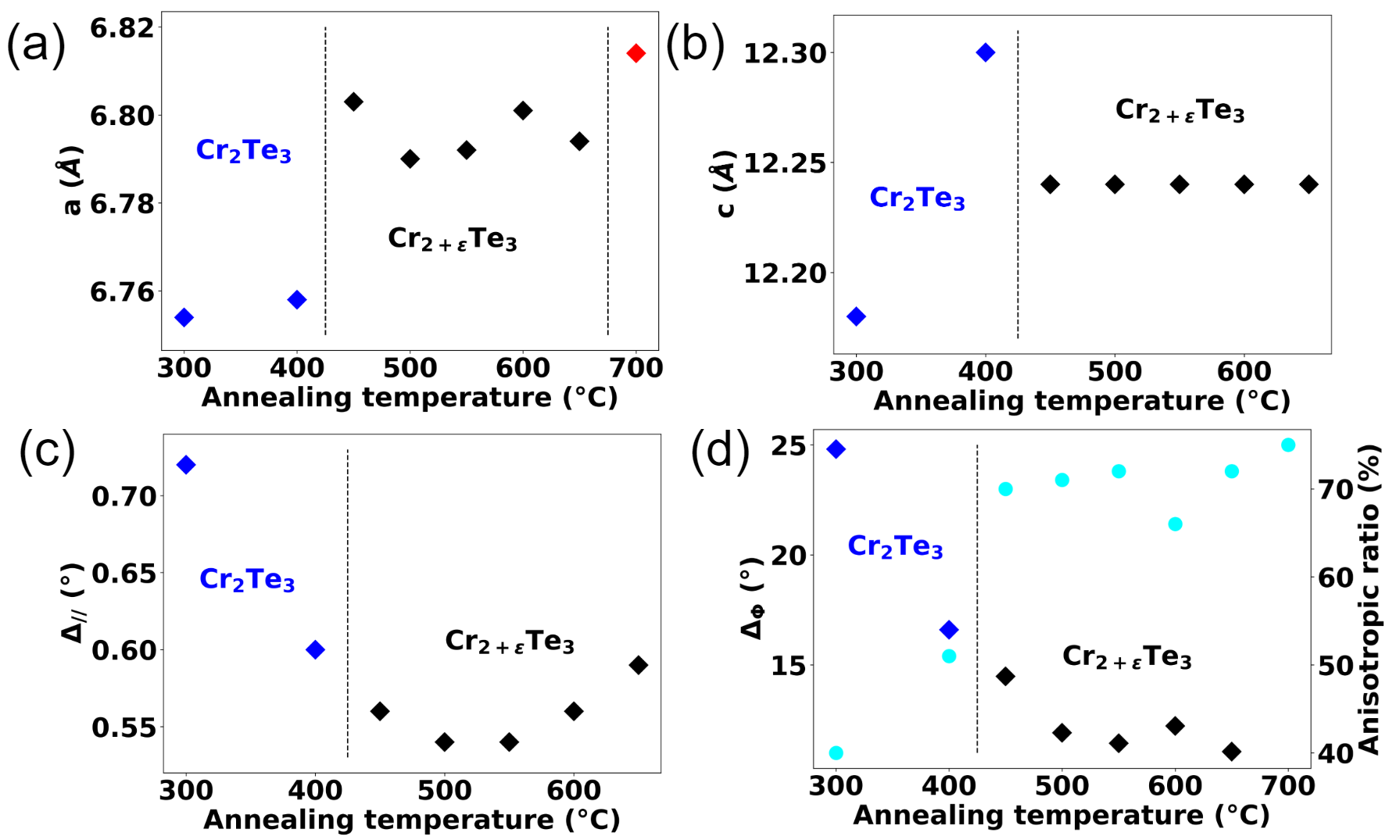}
    \caption{(a) Evolution of the in-plane lattice parameter $a$. The diamonds in blue indicate the Cr$_{1.33}$Te$_2$ phase, the diamonds in black indicate a phase with $x$>0.33 and in red a phase close to metallic Cr. (b) Evolution of the out-of plane lattice parameter $c$. (c) Evolution of the FWHM $\Delta_{\parallel}$ of the Cr$_{1.33}$Te$_2$ (300) XRD peak in the radial scan. (d) Left axis: evolution of the FWHM $\Delta_{\Phi}$ of the Cr$_{1.33}$Te$_2$ (300) XRD peak (diamonds) in the azimuthal scan. Right axis: evolution of the anisotropic ratio defined in Fig.~\ref{Fig2}d (light blue dots).
    }
\label{Fig3}
\end{center}
\end{figure*}
\begin{figure}[ht]
    \begin{center}
    \includegraphics[width=8.6cm]{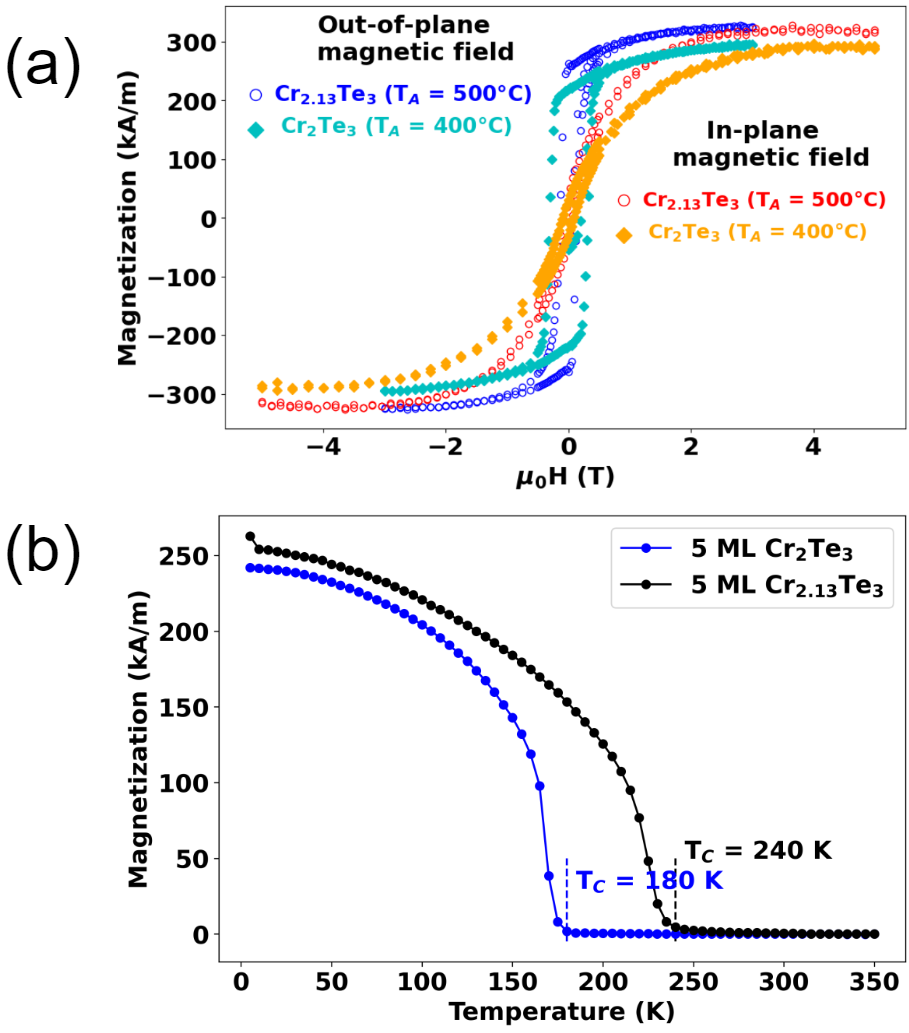}
    \caption{(a) Hysteresis loops obtained by SQUID for two phases of \CxT\ (open and full symbols). The signals were measured at 5 K with a magnetic field applied either perpendicularly or parallel to the layers. (b) Remanent magnetization curves of 5 ML of Cr$_{1.33}$Te$_2$ deposited on graphene/SiC measured with SQUID magnetometry without external field. The first sample (in blue) was annealed at 400°C and corresponds to the Cr$_{1.33}$Te$_2$ phase. The second one (in black) was annealed at 500°C with a final phase of Cr$_{1.375}$Te$_2$.}
\label{Fig4}
\end{center}
\end{figure}
\begin{figure*}[ht]
    \begin{center}
    \includegraphics[width=17.2cm]{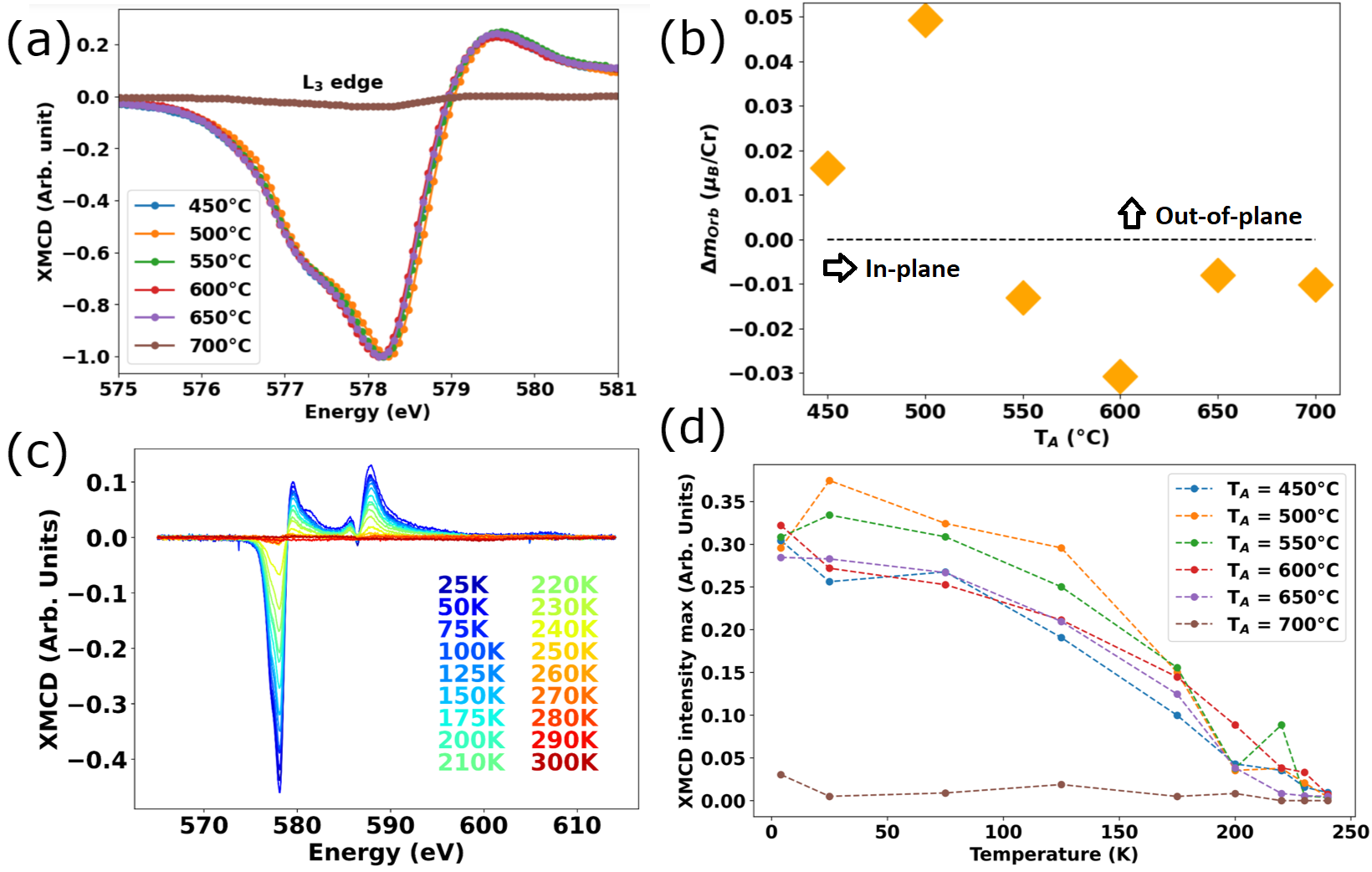}
    \caption{(a) Normalized XMCD spectra around the L$_3$ edge of Cr for 5 unit cell thick films of \CxT\ as a function of the annealing temperature. (b) Difference of Cr magnetic orbital momenta obtained using the XMCD sum rules between the normal and grazing (60° off the normal) incidence signals. (c) XMCD spectra of a Cr$_{1.33}$Te$_2$ sample annealed at 500°C as a function of temperature with a perpendicular magnetic field of 10 mT. (d) Remanent magnetization curves extracted from the XMCD spectra as a function of the annealing temperature.
    }
\label{Fig5}
\end{center}
\end{figure*}

\pagebreak
Two-dimensional (2D) ferromagnetic materials have opened new perspectives both in fundamental physics to explore low dimensionality magnetism \cite{Wang2022} but also in applications to build ultra-compact magnetic memories and sensors \cite{Yang_2022}. Two-dimensional ferromagnetism was first discovered in semiconducting CrI$_3$ \cite{Huang_2017} and Cr$_2$Ge$_2$Te$_6$ \cite{Gong_2017}. Since then, the number of 2D ferromagnets has been increasing but most studies have been performed on mechanically exfoliated flakes. The progress in molecular beam epitaxy (MBE) gives today the opportunity to grow high crystalline quality 2D magnets on large areas \cite{Lv_2023, Lopes_2021,Kotsaki_2022,Ribeiro_2022,Velez_2022,Li_2023,guillet_epitaxial_2023}. Among them, the metallic Fe$_x$GeTe$_2$ ($x$=3-5) and \CxT\ ($x$=0-1) families are very promising since they exhibit the highest Curie temperatures up to room temperature. The magnetic properties of \CxT\ evolve with the chromium content $x$ from a Curie temperature (\Tc) close to room temperature ($\approx$315 K) and in-plane magnetic anisotropy for CrTe$_2$ \cite{freitas_ferromagnetism_2015,purbawati_-plane_2020}, to a \Tc\ of 180 K and perpendicular magnetic anisotropy (PMA) for Cr$_{1.33}$Te$_2$ (\CT) \cite{dijkstrat_band-structurecalculations_1989, guillet_epitaxial_2023} to again above room temperature (\Tc$\approx$340 K) and in-plane magnetic anisotropy for Cr$_{1.5}$Te$_2$ (Cr$_3$Te$_4$) \cite{lasek_van_2022, fujisawa_tailoring_2020}.
Several studies were performed to control the magnetic properties of these materials grown by MBE. Fujisawa et al. \cite{fujisawa_tailoring_2020} as well as Wang et al. \cite{wang_layer-number-independent_2022} both reported on the evolution of the magnetic properties of \CxT\ layers grown on sapphire with in situ post-growth annealing. The first study correlates the changes with the film composition, evolving from Cr$_{1.33}$Te$_2$ to Cr$_{1.82}$Te$_2$. In the second, the authors claim that the stoichiometry is constant and attribute the evolution to strain in the layers.\\
In this work, we study the structural and magnetic properties of thin Cr$_{1.33}$Te$_2$ (\CT) films grown on graphene as a function of the post-growth annealing temperature (T$_A$). Thin layers present several advantages in the perspective of integrating such 2D ferromagnets in ultra compact spintronic devices like magnetic random access memories (MRAMs). First, the efficiency of spin-transfer torque (STT) or spin-orbit torque (SOT) writing mechanisms is inversely proportional to the ferromagnetic layer thickness. Thus, the magnetization of thin layers will be easier to switch reducing then the critical current and power consumption. This makes 2D ferromagnets good candidates for such devices thanks to their ultimately low thickness. Second, to manipulate the magnetic anisotropy with a voltage (known as the Voltage-Controlled Magnetic Anisotropy or VCMA procedure) in magnetic memories, the ferromagnetic layer should be thin enough to limit electric field screening. 2D ferromagnets are mostly metallic but their two-dimensional character is a real asset to manipulate their magnetic properties (magnetic anisotropy or Curie temperature) with a voltage since the thickness is less than the screening length.
Moreover, the thin character of the layers and the vdW interaction with graphene\cite{Dau_2018,dappe_charge_2020,Abdukayumov_2024} should favor atomic mobility and then a drastic change of composition and structural properties with annealing. Here, we report a change of the composition for T$_A$ $\geq$ 450°C associated with a change of magnetic properties (both in terms of magnetic anisotropy and \Tc). However, the magnetic properties remain the same above a certain annealing temperature corresponding to a fully relaxed atomic structure on graphene with constant lattice parameters. Our findings demonstrate that the substrate plays a role in the evolution of the magnetic properties of \CxT\ compounds with post growth annealing which opens a new way for tailoring their magnetic properties.\\
The crystal structure of \CxT\ is formed by the stacking of CrTe$_2$ monolayers made of a plane of Cr atoms with valence number of II and III in between two Te planes and so-called intercalated Cr layers  (see Fig.~\ref{Fig1}(a)). These Cr planes are only partially filled and Cr atoms can have a different valence number (I in the case of Cr$_{1.33}$Te$_2$ corresponding to \CT\ for which 1/3 of the lattice sites of the intercalated planes are occupied).\\
We performed the MBE growth of 5 unit cells thick films of Cr$_{1.33}$Te$_2$ (\CT) ($\approx$6.1 nm) on  graphene/SiC \cite{guillet_epitaxial_2023} obtained by graphitization of 4H-SiC(0001) \cite{kumar_growth_2016}. The choice of graphene as a template layer for the epitaxial growth is motivated by the weak interaction with the ferromagnetic layer allowing a quasi free-standing character of Cr$_{1.33}$Te$_2$. Added to the thin character of the layers, we assumed this would reduce the energy cost of a composition change and facilitate the structural relaxation using thermal annealing. Moreover, it has a good thermal stability allowing to anneal the material without any degradation of the underlying structure. In order to change the composition of the thin films, annealing steps of 10 minutes under Te flux were performed at an annealing temperature T$_A$ ranging between 400°C and 700°C. Using Rutherford backscattering spectrometry (RBS), we measured the stoichiometry of the obtained layers after annealing (see supplementary material for details). The results are shown in Fig.~\ref{Fig1}(b). For T$_A$ $\leq$ 400°C, no composition change is observed. For 500°C $\leq$ T$_A$ $\leq$ 600°C, an increase of the Cr:Te ratio is measured with little variations within the error bars. For T$_A$ = 650°C, the Cr:Te ratio measured experimentally is close to 1 and for T$_A$ = 700°C, it is close to 18 with only traces of Te remaining. To sum up these results, the annealing steps above 500°C lead to an increase of the Cr:Te ratio until the full evaporation of Te atoms. This is in agreement with the study reported by Fujisawa et al. \cite{fujisawa_tailoring_2020} on 80 nm thick films of Cr$_{1.33}$Te$_2$ deposited on sapphire.\\
We then studied how these composition changes affected the lattice parameters and morphology of the layers. During the epitaxial growth and the annealing steps, the reflection high-energy electron diffraction (RHEED) technique was used to monitor the crystallinity of the layers surface. The diffraction pattern of 5 monolayers (ML) of Cr$_{1.33}$Te$_2$ after the growth on graphene/SiC at 300°C can be seen in Fig.~\ref{Fig2}(a). A streaky diffraction pattern can be observed indicating a flat and well-crystallized surface. Two azimuthal directions are displayed separated by an angle of 30°. The Cr$_{1.33}$Te$_2$ (100) diffraction rod can be observed along both directions (less intense along the [1$\bar{1}$0] direction) revealing that the crystal grains are only partially oriented along the graphene crystal directions. The RHEED patterns during growth show the epitaxial relationship between Cr$_2$Te$_3$ and graphene: Cr$_2$Te$_3$[100]//graphene[1$\bar{1}$0] (see Supplemental Information of Ref. \citenum{guillet_epitaxial_2023}). We did not observe any change in the position of the streaks during the growth at 300°C between 1 and 10 monolayers meaning that Cr$_2$Te$_3$ grows fully relaxed on graphene following the van der Waals epitaxy mechanism. Fig.~\ref{Fig2}(b) reveals the RHEED diffraction pattern of \CxT\ layers along the [100] crystal axis of graphene after the annealing step at different temperatures. The width of the diffraction rods is systematically reduced with respect to the one of as-grown layers at 300°C. For annealing temperatures between 400°C and 700°C, a (2$\times$2) surface reconstruction appears in the RHEED patterns. For the layers annealed at 700°C, other rods are visible (highlighted with blue arrows) that match the ones of graphene. This means that the \CxT\ layers either re-evaporated due to the annealing step or dewetted the surface. In order to gain more insight into the crystal structure of the layers, a capping layer of 3 nm of Al (that transforms into a passivation AlO$_x$ layer in air) was deposited before taking the samples out of the MBE chamber. Ex-situ characterizations were then performed such as out-of-plane specular and in-plane radial and azimuthal XRD scans (see supplementary material for details).
An example of in-plane radial scans is shown in Fig.~\ref{Fig2}(c) for layers annealed at 450°C. The diffraction spectra are measured along three azimuthal directions: the two-main crystallographic axis of graphene $\Phi$ = 0° ([100] direction) and $\Phi$ = 30° ([110] direction) as well as a direction in between ($\Phi$ = 15°) to obtain information about the crystal orientation. Only weak peaks corresponding to Cr$_{1.33}$Te$_2$ can be observed along the $\Phi$ = 15° direction (the intensity is shown in logarithmic scale) revealing that some grains have random orientations. Finally, the full-width at half maximum (FWHM, $\Delta_{\parallel}$) of the main Cr$_{1.33}$Te$_2$ diffraction peaks is measured to obtain information about the average grain size of the films as a function of the annealing temperature. An example of in-plane azimuthal scan for layers annealed at T$_A$ = 450°C is shown in Fig.~\ref{Fig2}(d). An intense peak is found around $\delta \phi$ = 30°, indicating that atomic coincidences between graphene and Cr$_{1.33}$Te$_2$ favor an orientation of the lattice along this direction. The unit cell of Cr$_{1.33}$Te$_2$ and graphene is rotated by 30° due to the larger periodicity induced by the intercalated Cr planes. The intensity profiles of in-plane radial scans around the Cr$_{1.33}$Te$_2$ (300) peak are displayed in grey at several azimuthal positions. They give the intensity baseline as well as a separation of the crystal grains in two categories. The first one is for the grains that are mostly oriented with a 30° angle with respect to the graphene [100] direction. It is highlighted in blue in Fig.~\ref{Fig2}(d). The rest of the crystal grains are randomly oriented and are shown in orange. This observation is in agreement with in-plane radial scans. The crystalline anisotropic ratio is then defined as the intensity ratio between those two grains categories: in Fig.~\ref{Fig2}(d), 70 \% of the crystal grains are in epitaxial relationship with graphene. For epitaxial grains, we obtain a mosaic spread ($\Delta_{\Phi}$) of 14.48° which is a common value for TMDs grown by van der Waals epitaxy on graphene \cite{Dau_2018}.
All the structural information obtained by XRD as a function of T$_A$ are summarized in Fig.~\ref{Fig3}. In order to highlight the composition change for the samples annealed at T$_A$ = 450°C or more, the data points corresponding to the structure of Cr$_{1.33}$Te$_2$ are indicated in blue compared to those in black for the annealed films (see Fig~\ref{Fig1}(c)). The layers annealed at 700°C exhibit a degraded crystal structure as shown by RHEED as a consequence of the evaporation of Te atoms measured by RBS. The in-plane lattice parameter of this sample is therefore indicated in red in Fig~\ref{Fig3}(a). A clear variation is observed with the composition change and the amount of intercalated Cr atoms. However, no significant variation is observed between T$_A$ = 450°C and 650°C. Similarly, for the out-of-plane lattice parameter, the same value was measured for all the annealed samples (Fig.~\ref{Fig3}(b)). This is in contradiction with the reported results by Fujisawa et al. \cite{fujisawa_tailoring_2020}, where a gradual increase of both lattice parameters was measured as a function of the increasing Cr:Te ratio. We argue that the difference is explained by the thinner character of our films (6.1 nm vs 80 nm) and the weaker interaction between thin \CxT\ films and graphene compared to sapphire resulting in an overall lattice relaxation which is not evolving until the final degradation at 700°C with an out-of-plane lattice parameter $c$ of 13.6 \AA, much larger than for the other samples. In Fig~\ref{Fig3}(c-d), the FWHM of the Cr$_{1.33}$Te$_2$ (300) diffraction peak for the in-plane radial and in-plane azimuthal scans respectively are displayed. They both reflect the crystalline quality of the films. In the first graph, a gradual decrease of $\Delta_{\parallel}$ with temperature is observed until T$_A$ = 550°C, followed by an increase. This result suggests that the post-growth annealing tends to increase the average size of the crystal grains until this temperature, but it seems that disorder starts appearing for annealing temperatures greater than 600°C, as clearly evidenced on the RHEED pattern for the sample annealed at 700°C. Fig.~\ref{Fig3}(d) shows that the annealing step favors the orientation of the \CxT\ crystal grains along the crystal axis of graphene, as both $\Delta\Phi$ is reduced (at least until 500°C) and the anisotropic ratio is increased. To conclude, most of the structural changes occur with the composition change from the initial Cr$_{1.33}$Te$_2$ phase for annealing temperatures above 450°C.\\
In order to correlate these measurements with magnetic properties, superconducting quantum interference device (SQUID) magnetometry was performed with the magnetic field applied parallel or perpendicular to the film plane. Fig.~\ref{Fig4}(a) shows magnetic hysteresis loops measured at 5 K for samples annealed at 400°C (full symbols, with the composition of Cr$_{1.33}$Te$_2$) and 500°C (empty symbols, Cr$_{1.375}$Te$_2$). In both cases, perpendicular magnetic anisotropy (PMA) is obtained with nevertheless a reduction of the magnetic anisotropy energy from 0.171 MJ/m$^3$ to 0.139 MJ/m$^3$ associated with the change of stoichiometry. Remanent magnetization measurements were also performed as can be seen in Fig.~\ref{Fig4}(b) by heating the samples in the absence of magnetic field after saturation with a perpendicular field of 3 Tesla. The Curie temperature \Tc\ increases from 180 K which is the typical value for bulk Cr$_{1.33}$Te$_2$ to 240 K. This demonstrates the possibility of changing the ordering temperature of very thin films (5 unit cells thick) by adjusting the film composition.
Finally, x-ray magnetic circular dichroism (XMCD) was performed on the DEIMOS beamline \cite{ohresser2014deimos} of the synchrotron SOLEIL (Saint Aubin, France). Experimental details and data treatment are given in the supplementary material. By measuring the L$_2$ and L$_3$ absorption edges of Cr atoms, we first aim at studying the spectroscopic evolution of XMCD spectra as a function of the composition change. Indeed, Cr atoms exhibit a different valence number in the intercalated planes which should be visible in the spectra. We normalized the intensity of the L$_3$ peak in Fig.~\ref{Fig5}(a) (the small variations can be explained by different magnetization and magnetic anisotropy) for measurements obtained at 5 K with a perpendicular applied magnetic field of $\pm$ 3 Tesla. Except for the sample annealed at 700°C, which is mostly metallic Cr, all the other samples yield a very similar spectrum, probably revealing that this technique is not sensitive enough to detect the change of Cr content in the intercalated planes. Nevertheless, we used these measurements to compute the magnetic orbital momentum of Cr atoms and magneto-crystalline anisotropy using the XMCD sum rules \cite{thole_x-ray_1992} (see details in supplementary material). As can be observed in Fig.~\ref{Fig5}(b), samples with T$_A$ $\leq$ 500°C exhibit PMA, whereas the samples annealed at higher temperatures show an in-plane easy axis of magnetization. These results correlate well with the experimental findings of Fujisawa et al. \cite{fujisawa_tailoring_2020}.
To investigate the evolution of \Tc\ with the annealing temperature $T_A$, XMCD spectra were recorded with a perpendicular field of 10 mT as a function of temperature for different annealing temperatures $T_A$, as displayed in Fig.~\ref{Fig5}(c). For each temperature, the sample was first re-saturated with a perpendicular field of 3 Tesla. The maximum absolute value of the XMCD spectra were then plotted as a function of temperature to deduce the \Tc\  (see Fig.~\ref{Fig5}(d)). Curie temperatures of the order of 250 K are found regardless the annealing temperature. This result seems in contradiction with the previous study on annealed Cr$_{1.33}$Te$_2$ films \cite{fujisawa_tailoring_2020}, where a gradual increase of the \Tc\ was reported from 180 K to 350 K. In comparison with the work of Fujisawa et al. \cite{fujisawa_tailoring_2020} where they observe a continuous increase of the Cr content and crystal expansion (both $a$ and $c$ lattice parameters increase) with the annealing temperature, we rather find a step-like behavior with a transition to a given Cr content and lattice parameters that remain constant up to the film degradation for T$_A$ $\geq$650°C. The two main differences are the film thickness (80 nm vs. 6.1 nm) and the substrate (sapphire vs. graphene). Therefore, it appears that the thermal treatment on sapphire with thicker film and strong interaction between the layer and the substrate favors the phase change of \CxT\ leading to a broad range of compositions, lattice parameters, magnetic anisotropy and \Tc\ from 180 K to 350 K. On the other hand, the weak vdW interaction between a thin film and graphene limits the action of thermal treatment to the formation of a single metastable phase with a given Cr content, lattice structure and a \Tc\ of 240-250 K. The magnetic anisotropy evolves more progressively from out-of-plane for T$_A$ $\leq$ 500°C to in-plane for T$_A$>500°C probably due to the reorganization of Cr atoms within the film. In order to elucidate the underlying mechanisms explaining these two different behaviors, further work is necessary in particular to model the dynamics of Cr atoms in the films upon annealing for a given film thickness and substrate.\\
In summary, this study brings complementary light to the control of the magnetic properties of \CxT\ compounds. Upon post growth annealing, Cr$_{1.33}$Te$_2$ (\CT) epitaxially grown on graphene with PMA and a \Tc\ of 180 K transforms into a phase of composition close to Cr$_{1.39}$Te$_2$ exhibiting in-plane magnetic anisotropy and a \Tc\ of 240-250 K. This phase remains stable upon further annealing at higher temperatures until it forms pure Cr when Te atoms evaporate above 650°C. This study demonstrates the complex interplay between composition, strain and magnetic properties in
\CxT\ compounds. They represent promising materials to be integrated in magnetic vdW heterostructures and spintronics for their adjustable magnetic properties.\\

\section*{Supplementary Material}
See online supplementary material for details about the Rutherford Backscattering method, x-ray diffraction method and x-ray magnetic circular dichroism measurements and data analysis. 

\begin{acknowledgments}
This work received funding from the European Union's Horizon 2020 research and innovation programme under grant agreement No 881603 (Graphene Flagship). The French National Research Agency (ANR) is acknowledged
for its support through the ANR-18-CE24-0007 MAGICVALLEY, ANR-20-CE24-0015 ELMAX
and ESR/EQUIPEX+ ANR-21-ESRE-0025 2D-MAG projects. The Grenoble Alpes University is acknowledged for the IDEX IRS/EVASPIN project. The LANEF framework (No. ANR-10-LABX-0051) is acknowledged for its support with mutualized infrastructure. XMCD experiments were performed on the DEIMOS beamline at SOLEIL Synchrotron, France (proposal number 20220542). We are grateful to the SOLEIL staff for smoothly running the facility.
\end{acknowledgments}

\section*{Data Availability Statement}
The data that support the findings of this study are available from the corresponding author upon reasonable request.

\bibliography{References.bib}

\begin{thebibliography}{23}%
\makeatletter
\providecommand \@ifxundefined [1]{%
 \@ifx{#1\undefined}
}%
\providecommand \@ifnum [1]{%
 \ifnum #1\expandafter \@firstoftwo
 \else \expandafter \@secondoftwo
 \fi
}%
\providecommand \@ifx [1]{%
 \ifx #1\expandafter \@firstoftwo
 \else \expandafter \@secondoftwo
 \fi
}%
\providecommand \natexlab [1]{#1}%
\providecommand \enquote  [1]{``#1''}%
\providecommand \bibnamefont  [1]{#1}%
\providecommand \bibfnamefont [1]{#1}%
\providecommand \citenamefont [1]{#1}%
\providecommand \href@noop [0]{\@secondoftwo}%
\providecommand \href [0]{\begingroup \@sanitize@url \@href}%
\providecommand \@href[1]{\@@startlink{#1}\@@href}%
\providecommand \@@href[1]{\endgroup#1\@@endlink}%
\providecommand \@sanitize@url [0]{\catcode `\\12\catcode `\$12\catcode
  `\&12\catcode `\#12\catcode `\^12\catcode `\_12\catcode `\%12\relax}%
\providecommand \@@startlink[1]{}%
\providecommand \@@endlink[0]{}%
\providecommand \url  [0]{\begingroup\@sanitize@url \@url }%
\providecommand \@url [1]{\endgroup\@href {#1}{\urlprefix }}%
\providecommand \urlprefix  [0]{URL }%
\providecommand \Eprint [0]{\href }%
\providecommand \doibase [0]{https://doi.org/}%
\providecommand \selectlanguage [0]{\@gobble}%
\providecommand \bibinfo  [0]{\@secondoftwo}%
\providecommand \bibfield  [0]{\@secondoftwo}%
\providecommand \translation [1]{[#1]}%
\providecommand \BibitemOpen [0]{}%
\providecommand \bibitemStop [0]{}%
\providecommand \bibitemNoStop [0]{.\EOS\space}%
\providecommand \EOS [0]{\spacefactor3000\relax}%
\providecommand \BibitemShut  [1]{\csname bibitem#1\endcsname}%
\let\auto@bib@innerbib\@empty
\bibitem [{\citenamefont {Wang}\ \emph
  {et~al.}(2022{\natexlab{a}})\citenamefont {Wang}, \citenamefont
  {Bedoya-Pinto}, \citenamefont {Blei}, \citenamefont {Dismukes}, \citenamefont
  {Hamo}, \citenamefont {Jenkins}, \citenamefont {Koperski}, \citenamefont
  {Liu}, \citenamefont {Sun}, \citenamefont {Telford}, \citenamefont {Kim},
  \citenamefont {Augustin}, \citenamefont {Vool}, \citenamefont {Yin},
  \citenamefont {Li}, \citenamefont {Falin}, \citenamefont {Dean},
  \citenamefont {Casanova}, \citenamefont {Evans}, \citenamefont {Chshiev},
  \citenamefont {Mishchenko}, \citenamefont {Petrovic}, \citenamefont {He},
  \citenamefont {Zhao}, \citenamefont {Tsen}, \citenamefont {Gerardot},
  \citenamefont {Brotons-Gisbert}, \citenamefont {Guguchia}, \citenamefont
  {Roy}, \citenamefont {Tongay}, \citenamefont {Wang}, \citenamefont {Hasan},
  \citenamefont {Wrachtrup}, \citenamefont {Yacoby}, \citenamefont {Fert},
  \citenamefont {Parkin}, \citenamefont {Novoselov}, \citenamefont {Dai},
  \citenamefont {Balicas},\ and\ \citenamefont {Santos}}]{Wang2022}%
  \BibitemOpen
  \bibfield  {author} {\bibinfo {author} {\bibfnamefont {Q.~H.}\ \bibnamefont
  {Wang}}, \bibinfo {author} {\bibfnamefont {A.}~\bibnamefont {Bedoya-Pinto}},
  \bibinfo {author} {\bibfnamefont {M.}~\bibnamefont {Blei}}, \bibinfo {author}
  {\bibfnamefont {A.~H.}\ \bibnamefont {Dismukes}}, \bibinfo {author}
  {\bibfnamefont {A.}~\bibnamefont {Hamo}}, \bibinfo {author} {\bibfnamefont
  {S.}~\bibnamefont {Jenkins}}, \bibinfo {author} {\bibfnamefont
  {M.}~\bibnamefont {Koperski}}, \bibinfo {author} {\bibfnamefont
  {Y.}~\bibnamefont {Liu}}, \bibinfo {author} {\bibfnamefont {Q.-C.}\
  \bibnamefont {Sun}}, \bibinfo {author} {\bibfnamefont {E.~J.}\ \bibnamefont
  {Telford}}, \bibinfo {author} {\bibfnamefont {H.~H.}\ \bibnamefont {Kim}},
  \bibinfo {author} {\bibfnamefont {M.}~\bibnamefont {Augustin}}, \bibinfo
  {author} {\bibfnamefont {U.}~\bibnamefont {Vool}}, \bibinfo {author}
  {\bibfnamefont {J.-X.}\ \bibnamefont {Yin}}, \bibinfo {author} {\bibfnamefont
  {L.~H.}\ \bibnamefont {Li}}, \bibinfo {author} {\bibfnamefont
  {A.}~\bibnamefont {Falin}}, \bibinfo {author} {\bibfnamefont {C.~R.}\
  \bibnamefont {Dean}}, \bibinfo {author} {\bibfnamefont {F.}~\bibnamefont
  {Casanova}}, \bibinfo {author} {\bibfnamefont {R.~F.~L.}\ \bibnamefont
  {Evans}}, \bibinfo {author} {\bibfnamefont {M.}~\bibnamefont {Chshiev}},
  \bibinfo {author} {\bibfnamefont {A.}~\bibnamefont {Mishchenko}}, \bibinfo
  {author} {\bibfnamefont {C.}~\bibnamefont {Petrovic}}, \bibinfo {author}
  {\bibfnamefont {R.}~\bibnamefont {He}}, \bibinfo {author} {\bibfnamefont
  {L.}~\bibnamefont {Zhao}}, \bibinfo {author} {\bibfnamefont {A.~W.}\
  \bibnamefont {Tsen}}, \bibinfo {author} {\bibfnamefont {B.~D.}\ \bibnamefont
  {Gerardot}}, \bibinfo {author} {\bibfnamefont {M.}~\bibnamefont
  {Brotons-Gisbert}}, \bibinfo {author} {\bibfnamefont {Z.}~\bibnamefont
  {Guguchia}}, \bibinfo {author} {\bibfnamefont {X.}~\bibnamefont {Roy}},
  \bibinfo {author} {\bibfnamefont {S.}~\bibnamefont {Tongay}}, \bibinfo
  {author} {\bibfnamefont {Z.}~\bibnamefont {Wang}}, \bibinfo {author}
  {\bibfnamefont {M.~Z.}\ \bibnamefont {Hasan}}, \bibinfo {author}
  {\bibfnamefont {J.}~\bibnamefont {Wrachtrup}}, \bibinfo {author}
  {\bibfnamefont {A.}~\bibnamefont {Yacoby}}, \bibinfo {author} {\bibfnamefont
  {A.}~\bibnamefont {Fert}}, \bibinfo {author} {\bibfnamefont {S.}~\bibnamefont
  {Parkin}}, \bibinfo {author} {\bibfnamefont {K.~S.}\ \bibnamefont
  {Novoselov}}, \bibinfo {author} {\bibfnamefont {P.}~\bibnamefont {Dai}},
  \bibinfo {author} {\bibfnamefont {L.}~\bibnamefont {Balicas}},\ and\ \bibinfo
  {author} {\bibfnamefont {E.~J.~G.}\ \bibnamefont {Santos}},\ }\bibfield
  {title} {\enquote {\bibinfo {title} {The magnetic genome of two-dimensional
  van der waals materials},}\ }\href {https://doi.org/10.1021/acsnano.1c09150}
  {\bibfield  {journal} {\bibinfo  {journal} {ACS Nano}\ }\textbf {\bibinfo
  {volume} {16}},\ \bibinfo {pages} {6960--7079} (\bibinfo {year}
  {2022}{\natexlab{a}})},\ \bibinfo {note} {pMID: 35442017}\BibitemShut
  {NoStop}%
\bibitem [{\citenamefont {Yang}\ \emph {et~al.}(2022)\citenamefont {Yang},
  \citenamefont {Valenzuela}, \citenamefont {Chshiev}, \citenamefont {Couet},
  \citenamefont {Dieny}, \citenamefont {Dlubak}, \citenamefont {Fert},
  \citenamefont {Garello}, \citenamefont {Jamet}, \citenamefont {Jeong},
  \citenamefont {Lee}, \citenamefont {Lee}, \citenamefont {Martin},
  \citenamefont {Kar}, \citenamefont {Sénéor}, \citenamefont {Shin},\ and\
  \citenamefont {Roche}}]{Yang_2022}%
  \BibitemOpen
  \bibfield  {author} {\bibinfo {author} {\bibfnamefont {H.}~\bibnamefont
  {Yang}}, \bibinfo {author} {\bibfnamefont {S.~O.}\ \bibnamefont
  {Valenzuela}}, \bibinfo {author} {\bibfnamefont {M.}~\bibnamefont {Chshiev}},
  \bibinfo {author} {\bibfnamefont {S.}~\bibnamefont {Couet}}, \bibinfo
  {author} {\bibfnamefont {B.}~\bibnamefont {Dieny}}, \bibinfo {author}
  {\bibfnamefont {B.}~\bibnamefont {Dlubak}}, \bibinfo {author} {\bibfnamefont
  {A.}~\bibnamefont {Fert}}, \bibinfo {author} {\bibfnamefont {K.}~\bibnamefont
  {Garello}}, \bibinfo {author} {\bibfnamefont {M.}~\bibnamefont {Jamet}},
  \bibinfo {author} {\bibfnamefont {D.-E.}\ \bibnamefont {Jeong}}, \bibinfo
  {author} {\bibfnamefont {K.}~\bibnamefont {Lee}}, \bibinfo {author}
  {\bibfnamefont {T.}~\bibnamefont {Lee}}, \bibinfo {author} {\bibfnamefont
  {M.-B.}\ \bibnamefont {Martin}}, \bibinfo {author} {\bibfnamefont {G.~S.}\
  \bibnamefont {Kar}}, \bibinfo {author} {\bibfnamefont {P.}~\bibnamefont
  {Sénéor}}, \bibinfo {author} {\bibfnamefont {H.-J.}\ \bibnamefont {Shin}},\
  and\ \bibinfo {author} {\bibfnamefont {S.}~\bibnamefont {Roche}},\ }\bibfield
   {title} {\enquote {\bibinfo {title} {Two-dimensional materials prospects for
  non-volatile spintronic memories},}\ }\href@noop {} {\bibfield  {journal}
  {\bibinfo  {journal} {Nature}\ }\textbf {\bibinfo {volume} {606}},\ \bibinfo
  {pages} {663--673} (\bibinfo {year} {2022})}\BibitemShut {NoStop}%
\bibitem [{\citenamefont {Huang}\ \emph {et~al.}(2017)\citenamefont {Huang},
  \citenamefont {Clark}, \citenamefont {Navarro-Moratalla}, \citenamefont
  {Klein}, \citenamefont {Cheng}, \citenamefont {Seyler}, \citenamefont
  {Zhong}, \citenamefont {Schmidgall}, \citenamefont {McGuire}, \citenamefont
  {Cobden}, \citenamefont {Yao}, \citenamefont {Xiao}, \citenamefont
  {Jarillo-Herrero},\ and\ \citenamefont {Xu}}]{Huang_2017}%
  \BibitemOpen
  \bibfield  {author} {\bibinfo {author} {\bibfnamefont {B.}~\bibnamefont
  {Huang}}, \bibinfo {author} {\bibfnamefont {G.}~\bibnamefont {Clark}},
  \bibinfo {author} {\bibfnamefont {E.}~\bibnamefont {Navarro-Moratalla}},
  \bibinfo {author} {\bibfnamefont {D.~R.}\ \bibnamefont {Klein}}, \bibinfo
  {author} {\bibfnamefont {R.}~\bibnamefont {Cheng}}, \bibinfo {author}
  {\bibfnamefont {K.~L.}\ \bibnamefont {Seyler}}, \bibinfo {author}
  {\bibfnamefont {D.}~\bibnamefont {Zhong}}, \bibinfo {author} {\bibfnamefont
  {E.}~\bibnamefont {Schmidgall}}, \bibinfo {author} {\bibfnamefont {M.~A.}\
  \bibnamefont {McGuire}}, \bibinfo {author} {\bibfnamefont {D.~H.}\
  \bibnamefont {Cobden}}, \bibinfo {author} {\bibfnamefont {W.}~\bibnamefont
  {Yao}}, \bibinfo {author} {\bibfnamefont {D.}~\bibnamefont {Xiao}}, \bibinfo
  {author} {\bibfnamefont {P.}~\bibnamefont {Jarillo-Herrero}},\ and\ \bibinfo
  {author} {\bibfnamefont {X.}~\bibnamefont {Xu}},\ }\bibfield  {title}
  {\enquote {\bibinfo {title} {Layer-dependent ferromagnetism in a van der
  waals crystal down to the monolayer limit},}\ }\href@noop {} {\bibfield
  {journal} {\bibinfo  {journal} {Nature}\ }\textbf {\bibinfo {volume} {546}},\
  \bibinfo {pages} {270--273} (\bibinfo {year} {2017})}\BibitemShut {NoStop}%
\bibitem [{\citenamefont {Gong}\ \emph {et~al.}(2017)\citenamefont {Gong},
  \citenamefont {Li}, \citenamefont {Li}, \citenamefont {Ji}, \citenamefont
  {Stern}, \citenamefont {Xia}, \citenamefont {Cao}, \citenamefont {Bao},
  \citenamefont {Wang}, \citenamefont {Wang}, \citenamefont {Qiu},
  \citenamefont {Cava}, \citenamefont {Louie}, \citenamefont {Xia},\ and\
  \citenamefont {Zhang}}]{Gong_2017}%
  \BibitemOpen
  \bibfield  {author} {\bibinfo {author} {\bibfnamefont {C.}~\bibnamefont
  {Gong}}, \bibinfo {author} {\bibfnamefont {L.}~\bibnamefont {Li}}, \bibinfo
  {author} {\bibfnamefont {Z.}~\bibnamefont {Li}}, \bibinfo {author}
  {\bibfnamefont {H.}~\bibnamefont {Ji}}, \bibinfo {author} {\bibfnamefont
  {A.}~\bibnamefont {Stern}}, \bibinfo {author} {\bibfnamefont
  {Y.}~\bibnamefont {Xia}}, \bibinfo {author} {\bibfnamefont {T.}~\bibnamefont
  {Cao}}, \bibinfo {author} {\bibfnamefont {W.}~\bibnamefont {Bao}}, \bibinfo
  {author} {\bibfnamefont {C.}~\bibnamefont {Wang}}, \bibinfo {author}
  {\bibfnamefont {Y.}~\bibnamefont {Wang}}, \bibinfo {author} {\bibfnamefont
  {Z.~Q.}\ \bibnamefont {Qiu}}, \bibinfo {author} {\bibfnamefont {R.~J.}\
  \bibnamefont {Cava}}, \bibinfo {author} {\bibfnamefont {S.~G.}\ \bibnamefont
  {Louie}}, \bibinfo {author} {\bibfnamefont {J.}~\bibnamefont {Xia}},\ and\
  \bibinfo {author} {\bibfnamefont {X.}~\bibnamefont {Zhang}},\ }\bibfield
  {title} {\enquote {\bibinfo {title} {Discovery of intrinsic ferromagnetism in
  two-dimensional van der waals crystals},}\ }\href@noop {} {\bibfield
  {journal} {\bibinfo  {journal} {Nature}\ }\textbf {\bibinfo {volume} {546}},\
  \bibinfo {pages} {265--269} (\bibinfo {year} {2017})}\BibitemShut {NoStop}%
\bibitem [{\citenamefont {Lv}\ \emph {et~al.}(2023)\citenamefont {Lv},
  \citenamefont {da~Silva}, \citenamefont {Figueroa}, \citenamefont
  {Guillemard}, \citenamefont {Aguirre}, \citenamefont {Camosi}, \citenamefont
  {Aballe}, \citenamefont {Valvidares}, \citenamefont {Valenzuela},
  \citenamefont {Schubert}, \citenamefont {Schmidbauer}, \citenamefont
  {Herfort}, \citenamefont {Hanke}, \citenamefont {Trampert}, \citenamefont
  {Engel-Herbert}, \citenamefont {Ramsteiner},\ and\ \citenamefont
  {Lopes}}]{Lv_2023}%
  \BibitemOpen
  \bibfield  {author} {\bibinfo {author} {\bibfnamefont {H.}~\bibnamefont
  {Lv}}, \bibinfo {author} {\bibfnamefont {A.}~\bibnamefont {da~Silva}},
  \bibinfo {author} {\bibfnamefont {A.~I.}\ \bibnamefont {Figueroa}}, \bibinfo
  {author} {\bibfnamefont {C.}~\bibnamefont {Guillemard}}, \bibinfo {author}
  {\bibfnamefont {I.~F.}\ \bibnamefont {Aguirre}}, \bibinfo {author}
  {\bibfnamefont {L.}~\bibnamefont {Camosi}}, \bibinfo {author} {\bibfnamefont
  {L.}~\bibnamefont {Aballe}}, \bibinfo {author} {\bibfnamefont
  {M.}~\bibnamefont {Valvidares}}, \bibinfo {author} {\bibfnamefont {S.~O.}\
  \bibnamefont {Valenzuela}}, \bibinfo {author} {\bibfnamefont
  {J.}~\bibnamefont {Schubert}}, \bibinfo {author} {\bibfnamefont
  {M.}~\bibnamefont {Schmidbauer}}, \bibinfo {author} {\bibfnamefont
  {J.}~\bibnamefont {Herfort}}, \bibinfo {author} {\bibfnamefont
  {M.}~\bibnamefont {Hanke}}, \bibinfo {author} {\bibfnamefont
  {A.}~\bibnamefont {Trampert}}, \bibinfo {author} {\bibfnamefont
  {R.}~\bibnamefont {Engel-Herbert}}, \bibinfo {author} {\bibfnamefont
  {M.}~\bibnamefont {Ramsteiner}},\ and\ \bibinfo {author} {\bibfnamefont
  {J.~M.~J.}\ \bibnamefont {Lopes}},\ }\bibfield  {title} {\enquote {\bibinfo
  {title} {Large-area synthesis of ferromagnetic fe5−xgete2/graphene van der
  waals heterostructures with curie temperature above room temperature},}\
  }\href@noop {} {\bibfield  {journal} {\bibinfo  {journal} {Small}\ }\textbf
  {\bibinfo {volume} {19}},\ \bibinfo {pages} {2302387} (\bibinfo {year}
  {2023})}\BibitemShut {NoStop}%
\bibitem [{\citenamefont {Lopes}\ \emph {et~al.}(2021)\citenamefont {Lopes},
  \citenamefont {Czubak}, \citenamefont {Zallo}, \citenamefont {Figueroa},
  \citenamefont {Guillemard}, \citenamefont {Valvidares}, \citenamefont
  {Rubio-Zuazo}, \citenamefont {López-Sanchéz}, \citenamefont {Valenzuela},
  \citenamefont {Hanke},\ and\ \citenamefont {Ramsteiner}}]{Lopes_2021}%
  \BibitemOpen
  \bibfield  {author} {\bibinfo {author} {\bibfnamefont {J.~M.~J.}\
  \bibnamefont {Lopes}}, \bibinfo {author} {\bibfnamefont {D.}~\bibnamefont
  {Czubak}}, \bibinfo {author} {\bibfnamefont {E.}~\bibnamefont {Zallo}},
  \bibinfo {author} {\bibfnamefont {A.~I.}\ \bibnamefont {Figueroa}}, \bibinfo
  {author} {\bibfnamefont {C.}~\bibnamefont {Guillemard}}, \bibinfo {author}
  {\bibfnamefont {M.}~\bibnamefont {Valvidares}}, \bibinfo {author}
  {\bibfnamefont {J.}~\bibnamefont {Rubio-Zuazo}}, \bibinfo {author}
  {\bibfnamefont {J.}~\bibnamefont {López-Sanchéz}}, \bibinfo {author}
  {\bibfnamefont {S.~O.}\ \bibnamefont {Valenzuela}}, \bibinfo {author}
  {\bibfnamefont {M.}~\bibnamefont {Hanke}},\ and\ \bibinfo {author}
  {\bibfnamefont {M.}~\bibnamefont {Ramsteiner}},\ }\bibfield  {title}
  {\enquote {\bibinfo {title} {Large-area van der waals epitaxy and magnetic
  characterization of fe3gete2 films on graphene},}\ }\href
  {https://doi.org/10.1088/2053-1583/ac171d} {\bibfield  {journal} {\bibinfo
  {journal} {2D Materials}\ }\textbf {\bibinfo {volume} {8}},\ \bibinfo {pages}
  {041001} (\bibinfo {year} {2021})}\BibitemShut {NoStop}%
\bibitem [{\citenamefont {Georgopoulou-Kotsaki}\ \emph
  {et~al.}(2023)\citenamefont {Georgopoulou-Kotsaki}, \citenamefont {Pappas},
  \citenamefont {Lintzeris}, \citenamefont {Tsipas}, \citenamefont {Fragkos},
  \citenamefont {Markou}, \citenamefont {Felser}, \citenamefont {Longo},
  \citenamefont {Fanciulli}, \citenamefont {Mantovan}, \citenamefont
  {Mahfouzi}, \citenamefont {Kioussis},\ and\ \citenamefont
  {Dimoulas}}]{Kotsaki_2022}%
  \BibitemOpen
  \bibfield  {author} {\bibinfo {author} {\bibfnamefont {E.}~\bibnamefont
  {Georgopoulou-Kotsaki}}, \bibinfo {author} {\bibfnamefont {P.}~\bibnamefont
  {Pappas}}, \bibinfo {author} {\bibfnamefont {A.}~\bibnamefont {Lintzeris}},
  \bibinfo {author} {\bibfnamefont {P.}~\bibnamefont {Tsipas}}, \bibinfo
  {author} {\bibfnamefont {S.}~\bibnamefont {Fragkos}}, \bibinfo {author}
  {\bibfnamefont {A.}~\bibnamefont {Markou}}, \bibinfo {author} {\bibfnamefont
  {C.}~\bibnamefont {Felser}}, \bibinfo {author} {\bibfnamefont
  {E.}~\bibnamefont {Longo}}, \bibinfo {author} {\bibfnamefont
  {M.}~\bibnamefont {Fanciulli}}, \bibinfo {author} {\bibfnamefont
  {R.}~\bibnamefont {Mantovan}}, \bibinfo {author} {\bibfnamefont
  {F.}~\bibnamefont {Mahfouzi}}, \bibinfo {author} {\bibfnamefont
  {N.}~\bibnamefont {Kioussis}},\ and\ \bibinfo {author} {\bibfnamefont
  {A.}~\bibnamefont {Dimoulas}},\ }\bibfield  {title} {\enquote {\bibinfo
  {title} {Significant enhancement of ferromagnetism above room temperature in
  epitaxial 2d van der waals ferromagnet fe5−δgete2/bi2te3
  heterostructures},}\ }\href@noop {} {\bibfield  {journal} {\bibinfo
  {journal} {Nanoscale}\ }\textbf {\bibinfo {volume} {15}},\ \bibinfo {pages}
  {2223--2233} (\bibinfo {year} {2023})}\BibitemShut {NoStop}%
\bibitem [{\citenamefont {Ribeiro}\ \emph {et~al.}(2022)\citenamefont
  {Ribeiro}, \citenamefont {Gentile}, \citenamefont {Marty}, \citenamefont
  {Dosenovic}, \citenamefont {Okuno}, \citenamefont {Vergnaud}, \citenamefont
  {Jacquot}, \citenamefont {Jalabert}, \citenamefont {Longo}, \citenamefont
  {Ohresser}, \citenamefont {Hallal}, \citenamefont {Chshiev}, \citenamefont
  {Boulle}, \citenamefont {Bonell},\ and\ \citenamefont
  {Jamet}}]{Ribeiro_2022}%
  \BibitemOpen
  \bibfield  {author} {\bibinfo {author} {\bibfnamefont {M.}~\bibnamefont
  {Ribeiro}}, \bibinfo {author} {\bibfnamefont {G.}~\bibnamefont {Gentile}},
  \bibinfo {author} {\bibfnamefont {A.}~\bibnamefont {Marty}}, \bibinfo
  {author} {\bibfnamefont {D.}~\bibnamefont {Dosenovic}}, \bibinfo {author}
  {\bibfnamefont {H.}~\bibnamefont {Okuno}}, \bibinfo {author} {\bibfnamefont
  {C.}~\bibnamefont {Vergnaud}}, \bibinfo {author} {\bibfnamefont {J.-F.}\
  \bibnamefont {Jacquot}}, \bibinfo {author} {\bibfnamefont {D.}~\bibnamefont
  {Jalabert}}, \bibinfo {author} {\bibfnamefont {D.}~\bibnamefont {Longo}},
  \bibinfo {author} {\bibfnamefont {P.}~\bibnamefont {Ohresser}}, \bibinfo
  {author} {\bibfnamefont {A.}~\bibnamefont {Hallal}}, \bibinfo {author}
  {\bibfnamefont {M.}~\bibnamefont {Chshiev}}, \bibinfo {author} {\bibfnamefont
  {O.}~\bibnamefont {Boulle}}, \bibinfo {author} {\bibfnamefont
  {F.}~\bibnamefont {Bonell}},\ and\ \bibinfo {author} {\bibfnamefont
  {M.}~\bibnamefont {Jamet}},\ }\bibfield  {title} {\enquote {\bibinfo {title}
  {Large-scale epitaxy of two-dimensional van der waals room-temperature
  ferromagnet {Fe5GeTe2}},}\ }\href@noop {} {\bibfield  {journal} {\bibinfo
  {journal} {npj 2D Materials and Applications}\ }\textbf {\bibinfo {volume}
  {6}},\ \bibinfo {pages} {10} (\bibinfo {year} {2022})}\BibitemShut {NoStop}%
\bibitem [{\citenamefont {Vélez-Fort}\ \emph {et~al.}(2022)\citenamefont
  {Vélez-Fort}, \citenamefont {Hallal}, \citenamefont {Sant}, \citenamefont
  {Guillet}, \citenamefont {Abdukayumov}, \citenamefont {Marty}, \citenamefont
  {Vergnaud}, \citenamefont {Jacquot}, \citenamefont {Jalabert}, \citenamefont
  {Fujii}, \citenamefont {Vobornik}, \citenamefont {Rault}, \citenamefont
  {Brookes}, \citenamefont {Longo}, \citenamefont {Ohresser}, \citenamefont
  {Ouerghi}, \citenamefont {Veuillen}, \citenamefont {Mallet}, \citenamefont
  {Boukari}, \citenamefont {Okuno}, \citenamefont {Chshiev}, \citenamefont
  {Bonell},\ and\ \citenamefont {Jamet}}]{Velez_2022}%
  \BibitemOpen
  \bibfield  {author} {\bibinfo {author} {\bibfnamefont {E.}~\bibnamefont
  {Vélez-Fort}}, \bibinfo {author} {\bibfnamefont {A.}~\bibnamefont {Hallal}},
  \bibinfo {author} {\bibfnamefont {R.}~\bibnamefont {Sant}}, \bibinfo {author}
  {\bibfnamefont {T.}~\bibnamefont {Guillet}}, \bibinfo {author} {\bibfnamefont
  {K.}~\bibnamefont {Abdukayumov}}, \bibinfo {author} {\bibfnamefont
  {A.}~\bibnamefont {Marty}}, \bibinfo {author} {\bibfnamefont
  {C.}~\bibnamefont {Vergnaud}}, \bibinfo {author} {\bibfnamefont {J.-F.}\
  \bibnamefont {Jacquot}}, \bibinfo {author} {\bibfnamefont {D.}~\bibnamefont
  {Jalabert}}, \bibinfo {author} {\bibfnamefont {J.}~\bibnamefont {Fujii}},
  \bibinfo {author} {\bibfnamefont {I.}~\bibnamefont {Vobornik}}, \bibinfo
  {author} {\bibfnamefont {J.}~\bibnamefont {Rault}}, \bibinfo {author}
  {\bibfnamefont {N.~B.}\ \bibnamefont {Brookes}}, \bibinfo {author}
  {\bibfnamefont {D.}~\bibnamefont {Longo}}, \bibinfo {author} {\bibfnamefont
  {P.}~\bibnamefont {Ohresser}}, \bibinfo {author} {\bibfnamefont
  {A.}~\bibnamefont {Ouerghi}}, \bibinfo {author} {\bibfnamefont {J.-Y.}\
  \bibnamefont {Veuillen}}, \bibinfo {author} {\bibfnamefont {P.}~\bibnamefont
  {Mallet}}, \bibinfo {author} {\bibfnamefont {H.}~\bibnamefont {Boukari}},
  \bibinfo {author} {\bibfnamefont {H.}~\bibnamefont {Okuno}}, \bibinfo
  {author} {\bibfnamefont {M.}~\bibnamefont {Chshiev}}, \bibinfo {author}
  {\bibfnamefont {F.}~\bibnamefont {Bonell}},\ and\ \bibinfo {author}
  {\bibfnamefont {M.}~\bibnamefont {Jamet}},\ }\bibfield  {title} {\enquote
  {\bibinfo {title} {Ferromagnetism and rashba spin–orbit coupling in the
  two-dimensional (v,pt)se2 alloy},}\ }\href@noop {} {\bibfield  {journal}
  {\bibinfo  {journal} {ACS Applied Electronic Materials}\ }\textbf {\bibinfo
  {volume} {4}},\ \bibinfo {pages} {259--268} (\bibinfo {year}
  {2022})}\BibitemShut {NoStop}%
\bibitem [{\citenamefont {Li}\ \emph {et~al.}(2023)\citenamefont {Li},
  \citenamefont {Liu}, \citenamefont {Sun}, \citenamefont {Zhu}, \citenamefont
  {Chen}, \citenamefont {Yang}, \citenamefont {Ai}, \citenamefont {Zhang},
  \citenamefont {Huang}, \citenamefont {Leng}, \citenamefont {Zhao},
  \citenamefont {Xie}, \citenamefont {Zhang}, \citenamefont {Joseph},
  \citenamefont {Banerjee}, \citenamefont {Narayan}, \citenamefont {Zou},
  \citenamefont {Liu}, \citenamefont {Xu},\ and\ \citenamefont
  {Xiu}}]{Li_2023}%
  \BibitemOpen
  \bibfield  {author} {\bibinfo {author} {\bibfnamefont {Z.}~\bibnamefont
  {Li}}, \bibinfo {author} {\bibfnamefont {S.}~\bibnamefont {Liu}}, \bibinfo
  {author} {\bibfnamefont {J.}~\bibnamefont {Sun}}, \bibinfo {author}
  {\bibfnamefont {J.}~\bibnamefont {Zhu}}, \bibinfo {author} {\bibfnamefont
  {Y.}~\bibnamefont {Chen}}, \bibinfo {author} {\bibfnamefont {Y.}~\bibnamefont
  {Yang}}, \bibinfo {author} {\bibfnamefont {L.}~\bibnamefont {Ai}}, \bibinfo
  {author} {\bibfnamefont {E.}~\bibnamefont {Zhang}}, \bibinfo {author}
  {\bibfnamefont {C.}~\bibnamefont {Huang}}, \bibinfo {author} {\bibfnamefont
  {P.}~\bibnamefont {Leng}}, \bibinfo {author} {\bibfnamefont {M.}~\bibnamefont
  {Zhao}}, \bibinfo {author} {\bibfnamefont {X.}~\bibnamefont {Xie}}, \bibinfo
  {author} {\bibfnamefont {Y.}~\bibnamefont {Zhang}}, \bibinfo {author}
  {\bibfnamefont {N.~B.}\ \bibnamefont {Joseph}}, \bibinfo {author}
  {\bibfnamefont {R.}~\bibnamefont {Banerjee}}, \bibinfo {author}
  {\bibfnamefont {A.}~\bibnamefont {Narayan}}, \bibinfo {author} {\bibfnamefont
  {J.}~\bibnamefont {Zou}}, \bibinfo {author} {\bibfnamefont {W.}~\bibnamefont
  {Liu}}, \bibinfo {author} {\bibfnamefont {X.}~\bibnamefont {Xu}},\ and\
  \bibinfo {author} {\bibfnamefont {F.}~\bibnamefont {Xiu}},\ }\bibfield
  {title} {\enquote {\bibinfo {title} {Low-pass filters based on van der waals
  ferromagnets},}\ }\href@noop {} {\bibfield  {journal} {\bibinfo  {journal}
  {Nature Electronics}\ }\textbf {\bibinfo {volume} {6}},\ \bibinfo {pages}
  {273--280} (\bibinfo {year} {2023})}\BibitemShut {NoStop}%
\bibitem [{\citenamefont {Guillet}\ \emph {et~al.}(2023)\citenamefont
  {Guillet}, \citenamefont {Vojáček}, \citenamefont {Dosenovic},
  \citenamefont {Ibrahim}, \citenamefont {Boukari}, \citenamefont {Li},
  \citenamefont {Choueikani}, \citenamefont {Ohresser}, \citenamefont
  {Ouerghi}, \citenamefont {Mesple}, \citenamefont {Renard}, \citenamefont
  {Jacquot}, \citenamefont {Jalabert}, \citenamefont {Okuno}, \citenamefont
  {Chshiev}, \citenamefont {Vergnaud}, \citenamefont {Bonell}, \citenamefont
  {Marty},\ and\ \citenamefont {Jamet}}]{guillet_epitaxial_2023}%
  \BibitemOpen
  \bibfield  {author} {\bibinfo {author} {\bibfnamefont {Q.}~\bibnamefont
  {Guillet}}, \bibinfo {author} {\bibfnamefont {L.}~\bibnamefont {Vojáček}},
  \bibinfo {author} {\bibfnamefont {D.}~\bibnamefont {Dosenovic}}, \bibinfo
  {author} {\bibfnamefont {F.}~\bibnamefont {Ibrahim}}, \bibinfo {author}
  {\bibfnamefont {H.}~\bibnamefont {Boukari}}, \bibinfo {author} {\bibfnamefont
  {J.}~\bibnamefont {Li}}, \bibinfo {author} {\bibfnamefont {F.}~\bibnamefont
  {Choueikani}}, \bibinfo {author} {\bibfnamefont {P.}~\bibnamefont
  {Ohresser}}, \bibinfo {author} {\bibfnamefont {A.}~\bibnamefont {Ouerghi}},
  \bibinfo {author} {\bibfnamefont {F.}~\bibnamefont {Mesple}}, \bibinfo
  {author} {\bibfnamefont {V.}~\bibnamefont {Renard}}, \bibinfo {author}
  {\bibfnamefont {J.-F.}\ \bibnamefont {Jacquot}}, \bibinfo {author}
  {\bibfnamefont {D.}~\bibnamefont {Jalabert}}, \bibinfo {author}
  {\bibfnamefont {H.}~\bibnamefont {Okuno}}, \bibinfo {author} {\bibfnamefont
  {M.}~\bibnamefont {Chshiev}}, \bibinfo {author} {\bibfnamefont
  {C.}~\bibnamefont {Vergnaud}}, \bibinfo {author} {\bibfnamefont
  {F.}~\bibnamefont {Bonell}}, \bibinfo {author} {\bibfnamefont
  {A.}~\bibnamefont {Marty}},\ and\ \bibinfo {author} {\bibfnamefont
  {M.}~\bibnamefont {Jamet}},\ }\bibfield  {title} {\enquote {\bibinfo {title}
  {Epitaxial van der {Waals} heterostructures of cr$_2$te$_3$ on
  two-dimensional materials},}\ }\href
  {https://doi.org/10.1103/PhysRevMaterials.7.054005} {\bibfield  {journal}
  {\bibinfo  {journal} {Physical Review Materials}\ }\textbf {\bibinfo {volume}
  {7}},\ \bibinfo {pages} {054005} (\bibinfo {year} {2023})}\BibitemShut
  {NoStop}%
\bibitem [{\citenamefont {Freitas}\ \emph {et~al.}(2015)\citenamefont
  {Freitas}, \citenamefont {Weht}, \citenamefont {Sulpice}, \citenamefont
  {Remenyi}, \citenamefont {Strobel}, \citenamefont {Gay}, \citenamefont
  {Marcus},\ and\ \citenamefont
  {Núñez-Regueiro}}]{freitas_ferromagnetism_2015}%
  \BibitemOpen
  \bibfield  {author} {\bibinfo {author} {\bibfnamefont {D.~C.}\ \bibnamefont
  {Freitas}}, \bibinfo {author} {\bibfnamefont {R.}~\bibnamefont {Weht}},
  \bibinfo {author} {\bibfnamefont {A.}~\bibnamefont {Sulpice}}, \bibinfo
  {author} {\bibfnamefont {G.}~\bibnamefont {Remenyi}}, \bibinfo {author}
  {\bibfnamefont {P.}~\bibnamefont {Strobel}}, \bibinfo {author} {\bibfnamefont
  {F.}~\bibnamefont {Gay}}, \bibinfo {author} {\bibfnamefont {J.}~\bibnamefont
  {Marcus}},\ and\ \bibinfo {author} {\bibfnamefont {M.}~\bibnamefont
  {Núñez-Regueiro}},\ }\bibfield  {title} {\enquote {\bibinfo {title}
  {Ferromagnetism in layered metastable 1 \textit{{T}} -{CrTe}
  $_{\textrm{2}}$},}\ }\href {https://doi.org/10.1088/0953-8984/27/17/176002}
  {\bibfield  {journal} {\bibinfo  {journal} {Journal of Physics: Condensed
  Matter}\ }\textbf {\bibinfo {volume} {27}},\ \bibinfo {pages} {176002}
  (\bibinfo {year} {2015})}\BibitemShut {NoStop}%
\bibitem [{\citenamefont {Purbawati}\ \emph {et~al.}(2020)\citenamefont
  {Purbawati}, \citenamefont {Coraux}, \citenamefont {Vogel}, \citenamefont
  {Hadj-Azzem}, \citenamefont {Wu}, \citenamefont {Bendiab}, \citenamefont
  {Jegouso}, \citenamefont {Renard}, \citenamefont {Marty}, \citenamefont
  {Bouchiat}, \citenamefont {Sulpice}, \citenamefont {Aballe}, \citenamefont
  {Foerster}, \citenamefont {Genuzio}, \citenamefont {Locatelli}, \citenamefont
  {Menteş}, \citenamefont {Han}, \citenamefont {Sun}, \citenamefont
  {Núñez-Regueiro},\ and\ \citenamefont
  {Rougemaille}}]{purbawati_-plane_2020}%
  \BibitemOpen
  \bibfield  {author} {\bibinfo {author} {\bibfnamefont {A.}~\bibnamefont
  {Purbawati}}, \bibinfo {author} {\bibfnamefont {J.}~\bibnamefont {Coraux}},
  \bibinfo {author} {\bibfnamefont {J.}~\bibnamefont {Vogel}}, \bibinfo
  {author} {\bibfnamefont {A.}~\bibnamefont {Hadj-Azzem}}, \bibinfo {author}
  {\bibfnamefont {N.}~\bibnamefont {Wu}}, \bibinfo {author} {\bibfnamefont
  {N.}~\bibnamefont {Bendiab}}, \bibinfo {author} {\bibfnamefont
  {D.}~\bibnamefont {Jegouso}}, \bibinfo {author} {\bibfnamefont
  {J.}~\bibnamefont {Renard}}, \bibinfo {author} {\bibfnamefont
  {L.}~\bibnamefont {Marty}}, \bibinfo {author} {\bibfnamefont
  {V.}~\bibnamefont {Bouchiat}}, \bibinfo {author} {\bibfnamefont
  {A.}~\bibnamefont {Sulpice}}, \bibinfo {author} {\bibfnamefont
  {L.}~\bibnamefont {Aballe}}, \bibinfo {author} {\bibfnamefont
  {M.}~\bibnamefont {Foerster}}, \bibinfo {author} {\bibfnamefont
  {F.}~\bibnamefont {Genuzio}}, \bibinfo {author} {\bibfnamefont
  {A.}~\bibnamefont {Locatelli}}, \bibinfo {author} {\bibfnamefont {T.~O.}\
  \bibnamefont {Menteş}}, \bibinfo {author} {\bibfnamefont {Z.~V.}\
  \bibnamefont {Han}}, \bibinfo {author} {\bibfnamefont {X.}~\bibnamefont
  {Sun}}, \bibinfo {author} {\bibfnamefont {M.}~\bibnamefont
  {Núñez-Regueiro}},\ and\ \bibinfo {author} {\bibfnamefont {N.}~\bibnamefont
  {Rougemaille}},\ }\bibfield  {title} {\enquote {\bibinfo {title} {In-{Plane}
  {Magnetic} {Domains} and {Néel}-like {Domain} {Walls} in {Thin} {Flakes} of
  the {Room} {Temperature} {CrTe2} {Van} der {Waals} {Ferromagnet}},}\ }\href
  {https://doi.org/10.1021/acsami.0c07017} {\bibfield  {journal} {\bibinfo
  {journal} {ACS Applied Materials \& Interfaces}\ }\textbf {\bibinfo {volume}
  {12}},\ \bibinfo {pages} {30702--30710} (\bibinfo {year} {2020})}\BibitemShut
  {NoStop}%
\bibitem [{\citenamefont {Dijkstrat}\ \emph {et~al.}(1989)\citenamefont
  {Dijkstrat}, \citenamefont {Weitering}, \citenamefont {van Bruggen},
  \citenamefont {Haast},\ and\ \citenamefont
  {de~Groot}}]{dijkstrat_band-structurecalculations_1989}%
  \BibitemOpen
  \bibfield  {author} {\bibinfo {author} {\bibfnamefont {J.}~\bibnamefont
  {Dijkstrat}}, \bibinfo {author} {\bibfnamefont {H.~H.}\ \bibnamefont
  {Weitering}}, \bibinfo {author} {\bibfnamefont {C.~F.}\ \bibnamefont {van
  Bruggen}}, \bibinfo {author} {\bibfnamefont {C.}~\bibnamefont {Haast}},\ and\
  \bibinfo {author} {\bibfnamefont {R.~A.}\ \bibnamefont {de~Groot}},\
  }\bibfield  {title} {\enquote {\bibinfo {title} {Band-structurecalculations,
  and magnetic and transport properties of ferromagnetic chromium tellurides
  ({CrTe}, {Cr3Te4},{Cr},{Te},)},}\ }\href@noop {} {\bibfield  {journal}
  {\bibinfo  {journal} {Journal of Physics: Condensed Matter}\ }\textbf
  {\bibinfo {volume} {1}},\ \bibinfo {pages} {9141} (\bibinfo {year}
  {1989})}\BibitemShut {NoStop}%
\bibitem [{\citenamefont {Lasek}\ \emph {et~al.}(2022)\citenamefont {Lasek},
  \citenamefont {Coelho}, \citenamefont {Gargiani}, \citenamefont {Valvidares},
  \citenamefont {Mohseni}, \citenamefont {Meyerheim}, \citenamefont
  {Kostanovskiy}, \citenamefont {Zberecki},\ and\ \citenamefont
  {Batzill}}]{lasek_van_2022}%
  \BibitemOpen
  \bibfield  {author} {\bibinfo {author} {\bibfnamefont {K.}~\bibnamefont
  {Lasek}}, \bibinfo {author} {\bibfnamefont {P.~M.}\ \bibnamefont {Coelho}},
  \bibinfo {author} {\bibfnamefont {P.}~\bibnamefont {Gargiani}}, \bibinfo
  {author} {\bibfnamefont {M.}~\bibnamefont {Valvidares}}, \bibinfo {author}
  {\bibfnamefont {K.}~\bibnamefont {Mohseni}}, \bibinfo {author} {\bibfnamefont
  {H.~L.}\ \bibnamefont {Meyerheim}}, \bibinfo {author} {\bibfnamefont
  {I.}~\bibnamefont {Kostanovskiy}}, \bibinfo {author} {\bibfnamefont
  {K.}~\bibnamefont {Zberecki}},\ and\ \bibinfo {author} {\bibfnamefont
  {M.}~\bibnamefont {Batzill}},\ }\bibfield  {title} {\enquote {\bibinfo
  {title} {Van der {Waals} epitaxy growth of {2D} ferromagnetic {Cr}(1+δ){Te2}
  nanolayers with concentration-tunable magnetic anisotropy},}\ }\href
  {https://doi.org/10.1063/5.0070079} {\bibfield  {journal} {\bibinfo
  {journal} {Applied Physics Reviews}\ }\textbf {\bibinfo {volume} {9}},\
  \bibinfo {pages} {011409} (\bibinfo {year} {2022})}\BibitemShut {NoStop}%
\bibitem [{\citenamefont {Fujisawa}\ \emph {et~al.}(2020)\citenamefont
  {Fujisawa}, \citenamefont {Pardo-Almanza}, \citenamefont {Garland},
  \citenamefont {Yamagami}, \citenamefont {Zhu}, \citenamefont {Chen},
  \citenamefont {Araki}, \citenamefont {Takeda}, \citenamefont {Kobayashi},
  \citenamefont {Takeda}, \citenamefont {Hsu}, \citenamefont {Chuang},
  \citenamefont {Laskowski}, \citenamefont {Khoo}, \citenamefont
  {Soumyanarayanan},\ and\ \citenamefont {Okada}}]{fujisawa_tailoring_2020}%
  \BibitemOpen
  \bibfield  {author} {\bibinfo {author} {\bibfnamefont {Y.}~\bibnamefont
  {Fujisawa}}, \bibinfo {author} {\bibfnamefont {M.}~\bibnamefont
  {Pardo-Almanza}}, \bibinfo {author} {\bibfnamefont {J.}~\bibnamefont
  {Garland}}, \bibinfo {author} {\bibfnamefont {K.}~\bibnamefont {Yamagami}},
  \bibinfo {author} {\bibfnamefont {X.}~\bibnamefont {Zhu}}, \bibinfo {author}
  {\bibfnamefont {X.}~\bibnamefont {Chen}}, \bibinfo {author} {\bibfnamefont
  {K.}~\bibnamefont {Araki}}, \bibinfo {author} {\bibfnamefont
  {T.}~\bibnamefont {Takeda}}, \bibinfo {author} {\bibfnamefont
  {M.}~\bibnamefont {Kobayashi}}, \bibinfo {author} {\bibfnamefont
  {Y.}~\bibnamefont {Takeda}}, \bibinfo {author} {\bibfnamefont {C.~H.}\
  \bibnamefont {Hsu}}, \bibinfo {author} {\bibfnamefont {F.~C.}\ \bibnamefont
  {Chuang}}, \bibinfo {author} {\bibfnamefont {R.}~\bibnamefont {Laskowski}},
  \bibinfo {author} {\bibfnamefont {K.~H.}\ \bibnamefont {Khoo}}, \bibinfo
  {author} {\bibfnamefont {A.}~\bibnamefont {Soumyanarayanan}},\ and\ \bibinfo
  {author} {\bibfnamefont {Y.}~\bibnamefont {Okada}},\ }\bibfield  {title}
  {\enquote {\bibinfo {title} {Tailoring magnetism in self-intercalated
  {Cr1}+{xTe} 2 epitaxial films},}\ }\href
  {https://doi.org/10.1103/PhysRevMaterials.4.114001} {\bibfield  {journal}
  {\bibinfo  {journal} {Physical Review Materials}\ }\textbf {\bibinfo {volume}
  {4}},\ \bibinfo {pages} {114001} (\bibinfo {year} {2020})}\BibitemShut
  {NoStop}%
\bibitem [{\citenamefont {Wang}\ \emph
  {et~al.}(2022{\natexlab{b}})\citenamefont {Wang}, \citenamefont {Kajihara},
  \citenamefont {Matsuoka}, \citenamefont {Saika}, \citenamefont {Yamagami},
  \citenamefont {Takeda}, \citenamefont {Wadati}, \citenamefont {Ishizaka},
  \citenamefont {Iwasa},\ and\ \citenamefont
  {Nakano}}]{wang_layer-number-independent_2022}%
  \BibitemOpen
  \bibfield  {author} {\bibinfo {author} {\bibfnamefont {Y.}~\bibnamefont
  {Wang}}, \bibinfo {author} {\bibfnamefont {S.}~\bibnamefont {Kajihara}},
  \bibinfo {author} {\bibfnamefont {H.}~\bibnamefont {Matsuoka}}, \bibinfo
  {author} {\bibfnamefont {B.~K.}\ \bibnamefont {Saika}}, \bibinfo {author}
  {\bibfnamefont {K.}~\bibnamefont {Yamagami}}, \bibinfo {author}
  {\bibfnamefont {Y.}~\bibnamefont {Takeda}}, \bibinfo {author} {\bibfnamefont
  {H.}~\bibnamefont {Wadati}}, \bibinfo {author} {\bibfnamefont
  {K.}~\bibnamefont {Ishizaka}}, \bibinfo {author} {\bibfnamefont
  {Y.}~\bibnamefont {Iwasa}},\ and\ \bibinfo {author} {\bibfnamefont
  {M.}~\bibnamefont {Nakano}},\ }\bibfield  {title} {\enquote {\bibinfo {title}
  {Layer-{Number}-{Independent} {Two}-{Dimensional} {Ferromagnetism} in
  {Cr3Te4}},}\ }\href {https://doi.org/10.1021/acs.nanolett.2c03532} {\bibfield
   {journal} {\bibinfo  {journal} {Nano Letters}\ }\textbf {\bibinfo {volume}
  {22}},\ \bibinfo {pages} {9964--9971} (\bibinfo {year}
  {2022}{\natexlab{b}})}\BibitemShut {NoStop}%
\bibitem [{\citenamefont {Dau}\ \emph {et~al.}(2018)\citenamefont {Dau},
  \citenamefont {Gay}, \citenamefont {Di~Felice}, \citenamefont {Vergnaud},
  \citenamefont {Marty}, \citenamefont {Beigné}, \citenamefont {Renaud},
  \citenamefont {Renault}, \citenamefont {Mallet}, \citenamefont {Le~Quang},
  \citenamefont {Veuillen}, \citenamefont {Huder}, \citenamefont {Renard},
  \citenamefont {Chapelier}, \citenamefont {Zamborlini}, \citenamefont
  {Jugovac}, \citenamefont {Feyer}, \citenamefont {Dappe}, \citenamefont
  {Pochet},\ and\ \citenamefont {Jamet}}]{Dau_2018}%
  \BibitemOpen
  \bibfield  {author} {\bibinfo {author} {\bibfnamefont {M.~T.}\ \bibnamefont
  {Dau}}, \bibinfo {author} {\bibfnamefont {M.}~\bibnamefont {Gay}}, \bibinfo
  {author} {\bibfnamefont {D.}~\bibnamefont {Di~Felice}}, \bibinfo {author}
  {\bibfnamefont {C.}~\bibnamefont {Vergnaud}}, \bibinfo {author}
  {\bibfnamefont {A.}~\bibnamefont {Marty}}, \bibinfo {author} {\bibfnamefont
  {C.}~\bibnamefont {Beigné}}, \bibinfo {author} {\bibfnamefont
  {G.}~\bibnamefont {Renaud}}, \bibinfo {author} {\bibfnamefont
  {O.}~\bibnamefont {Renault}}, \bibinfo {author} {\bibfnamefont
  {P.}~\bibnamefont {Mallet}}, \bibinfo {author} {\bibfnamefont
  {T.}~\bibnamefont {Le~Quang}}, \bibinfo {author} {\bibfnamefont {J.-Y.}\
  \bibnamefont {Veuillen}}, \bibinfo {author} {\bibfnamefont {L.}~\bibnamefont
  {Huder}}, \bibinfo {author} {\bibfnamefont {V.~T.}\ \bibnamefont {Renard}},
  \bibinfo {author} {\bibfnamefont {C.}~\bibnamefont {Chapelier}}, \bibinfo
  {author} {\bibfnamefont {G.}~\bibnamefont {Zamborlini}}, \bibinfo {author}
  {\bibfnamefont {M.}~\bibnamefont {Jugovac}}, \bibinfo {author} {\bibfnamefont
  {V.}~\bibnamefont {Feyer}}, \bibinfo {author} {\bibfnamefont {Y.~J.}\
  \bibnamefont {Dappe}}, \bibinfo {author} {\bibfnamefont {P.}~\bibnamefont
  {Pochet}},\ and\ \bibinfo {author} {\bibfnamefont {M.}~\bibnamefont
  {Jamet}},\ }\bibfield  {title} {\enquote {\bibinfo {title} {Beyond van der
  waals interaction: The case of mose2 epitaxially grown on few-layer
  graphene},}\ }\href {https://doi.org/10.1021/acsnano.7b07446} {\bibfield
  {journal} {\bibinfo  {journal} {ACS Nano}\ }\textbf {\bibinfo {volume}
  {12}},\ \bibinfo {pages} {2319--2331} (\bibinfo {year} {2018})},\ \bibinfo
  {note} {pMID: 29384649},\ \Eprint
  {https://arxiv.org/abs/https://doi.org/10.1021/acsnano.7b07446}
  {https://doi.org/10.1021/acsnano.7b07446} \BibitemShut {NoStop}%
\bibitem [{\citenamefont {Dappe}\ \emph {et~al.}(2020)\citenamefont {Dappe},
  \citenamefont {Almadori}, \citenamefont {Dau}, \citenamefont {Vergnaud},
  \citenamefont {Jamet}, \citenamefont {Paillet}, \citenamefont {Journot},
  \citenamefont {Hyot}, \citenamefont {Pochet},\ and\ \citenamefont
  {Grévin}}]{dappe_charge_2020}%
  \BibitemOpen
  \bibfield  {author} {\bibinfo {author} {\bibfnamefont {Y.~J.}\ \bibnamefont
  {Dappe}}, \bibinfo {author} {\bibfnamefont {Y.}~\bibnamefont {Almadori}},
  \bibinfo {author} {\bibfnamefont {M.~T.}\ \bibnamefont {Dau}}, \bibinfo
  {author} {\bibfnamefont {C.}~\bibnamefont {Vergnaud}}, \bibinfo {author}
  {\bibfnamefont {M.}~\bibnamefont {Jamet}}, \bibinfo {author} {\bibfnamefont
  {C.}~\bibnamefont {Paillet}}, \bibinfo {author} {\bibfnamefont
  {T.}~\bibnamefont {Journot}}, \bibinfo {author} {\bibfnamefont
  {B.}~\bibnamefont {Hyot}}, \bibinfo {author} {\bibfnamefont {P.}~\bibnamefont
  {Pochet}},\ and\ \bibinfo {author} {\bibfnamefont {B.}~\bibnamefont
  {Grévin}},\ }\bibfield  {title} {\enquote {\bibinfo {title} {Charge
  transfers and charged defects in {WSe2}/graphene-{SiC} interfaces},}\ }\href
  {https://doi.org/10.1088/1361-6528/ab8083} {\bibfield  {journal} {\bibinfo
  {journal} {Nanotechnology}\ }\textbf {\bibinfo {volume} {31}},\ \bibinfo
  {pages} {255709} (\bibinfo {year} {2020})}\BibitemShut {NoStop}%
\bibitem [{\citenamefont {Abdukayumov}\ \emph {et~al.}(2024)\citenamefont
  {Abdukayumov}, \citenamefont {Mičica}, \citenamefont {Ibrahim},
  \citenamefont {Vojáček}, \citenamefont {Vergnaud}, \citenamefont {Marty},
  \citenamefont {Veuillen}, \citenamefont {Mallet}, \citenamefont {de~Moraes},
  \citenamefont {Dosenovic}, \citenamefont {Gambarelli}, \citenamefont
  {Maurel}, \citenamefont {Wright}, \citenamefont {Tignon}, \citenamefont
  {Mangeney}, \citenamefont {Ouerghi}, \citenamefont {Renard}, \citenamefont
  {Mesple}, \citenamefont {Li}, \citenamefont {Bonell}, \citenamefont {Okuno},
  \citenamefont {Chshiev}, \citenamefont {George}, \citenamefont {Jaffrès},
  \citenamefont {Dhillon},\ and\ \citenamefont {Jamet}}]{Abdukayumov_2024}%
  \BibitemOpen
  \bibfield  {author} {\bibinfo {author} {\bibfnamefont {K.}~\bibnamefont
  {Abdukayumov}}, \bibinfo {author} {\bibfnamefont {M.}~\bibnamefont
  {Mičica}}, \bibinfo {author} {\bibfnamefont {F.}~\bibnamefont {Ibrahim}},
  \bibinfo {author} {\bibfnamefont {L.}~\bibnamefont {Vojáček}}, \bibinfo
  {author} {\bibfnamefont {C.}~\bibnamefont {Vergnaud}}, \bibinfo {author}
  {\bibfnamefont {A.}~\bibnamefont {Marty}}, \bibinfo {author} {\bibfnamefont
  {J.-Y.}\ \bibnamefont {Veuillen}}, \bibinfo {author} {\bibfnamefont
  {P.}~\bibnamefont {Mallet}}, \bibinfo {author} {\bibfnamefont {I.~G.}\
  \bibnamefont {de~Moraes}}, \bibinfo {author} {\bibfnamefont {D.}~\bibnamefont
  {Dosenovic}}, \bibinfo {author} {\bibfnamefont {S.}~\bibnamefont
  {Gambarelli}}, \bibinfo {author} {\bibfnamefont {V.}~\bibnamefont {Maurel}},
  \bibinfo {author} {\bibfnamefont {A.}~\bibnamefont {Wright}}, \bibinfo
  {author} {\bibfnamefont {J.}~\bibnamefont {Tignon}}, \bibinfo {author}
  {\bibfnamefont {J.}~\bibnamefont {Mangeney}}, \bibinfo {author}
  {\bibfnamefont {A.}~\bibnamefont {Ouerghi}}, \bibinfo {author} {\bibfnamefont
  {V.}~\bibnamefont {Renard}}, \bibinfo {author} {\bibfnamefont
  {F.}~\bibnamefont {Mesple}}, \bibinfo {author} {\bibfnamefont
  {J.}~\bibnamefont {Li}}, \bibinfo {author} {\bibfnamefont {F.}~\bibnamefont
  {Bonell}}, \bibinfo {author} {\bibfnamefont {H.}~\bibnamefont {Okuno}},
  \bibinfo {author} {\bibfnamefont {M.}~\bibnamefont {Chshiev}}, \bibinfo
  {author} {\bibfnamefont {J.-M.}\ \bibnamefont {George}}, \bibinfo {author}
  {\bibfnamefont {H.}~\bibnamefont {Jaffrès}}, \bibinfo {author}
  {\bibfnamefont {S.}~\bibnamefont {Dhillon}},\ and\ \bibinfo {author}
  {\bibfnamefont {M.}~\bibnamefont {Jamet}},\ }\bibfield  {title} {\enquote
  {\bibinfo {title} {Atomic-layer controlled transition from inverse
  rashba–edelstein effect to inverse spin hall effect in 2d ptse2 probed by
  thz spintronic emission},}\ }\href
  {https://doi.org/https://doi.org/10.1002/adma.202304243} {\bibfield
  {journal} {\bibinfo  {journal} {Advanced Materials}\ }\textbf {\bibinfo
  {volume} {36}},\ \bibinfo {pages} {2304243} (\bibinfo {year} {2024})},\
  \Eprint
  {https://arxiv.org/abs/https://onlinelibrary.wiley.com/doi/pdf/10.1002/adma.202304243}
  {https://onlinelibrary.wiley.com/doi/pdf/10.1002/adma.202304243} \BibitemShut
  {NoStop}%
\bibitem [{\citenamefont {Kumar}\ \emph {et~al.}(2016)\citenamefont {Kumar},
  \citenamefont {Baraket}, \citenamefont {Paillet}, \citenamefont {Huntzinger},
  \citenamefont {Tiberj}, \citenamefont {Jansen}, \citenamefont {Vila},
  \citenamefont {Cubuku}, \citenamefont {Vergnaud}, \citenamefont {Jamet},
  \citenamefont {Lapertot}, \citenamefont {Rouchon}, \citenamefont {Zahab},
  \citenamefont {Sauvajol}, \citenamefont {Dubois}, \citenamefont {Lefloch},\
  and\ \citenamefont {Duclairoir}}]{kumar_growth_2016}%
  \BibitemOpen
  \bibfield  {author} {\bibinfo {author} {\bibfnamefont {B.}~\bibnamefont
  {Kumar}}, \bibinfo {author} {\bibfnamefont {M.}~\bibnamefont {Baraket}},
  \bibinfo {author} {\bibfnamefont {M.}~\bibnamefont {Paillet}}, \bibinfo
  {author} {\bibfnamefont {J.-R.}\ \bibnamefont {Huntzinger}}, \bibinfo
  {author} {\bibfnamefont {A.}~\bibnamefont {Tiberj}}, \bibinfo {author}
  {\bibfnamefont {A.}~\bibnamefont {Jansen}}, \bibinfo {author} {\bibfnamefont
  {L.}~\bibnamefont {Vila}}, \bibinfo {author} {\bibfnamefont {M.}~\bibnamefont
  {Cubuku}}, \bibinfo {author} {\bibfnamefont {C.}~\bibnamefont {Vergnaud}},
  \bibinfo {author} {\bibfnamefont {M.}~\bibnamefont {Jamet}}, \bibinfo
  {author} {\bibfnamefont {G.}~\bibnamefont {Lapertot}}, \bibinfo {author}
  {\bibfnamefont {D.}~\bibnamefont {Rouchon}}, \bibinfo {author} {\bibfnamefont
  {A.-A.}\ \bibnamefont {Zahab}}, \bibinfo {author} {\bibfnamefont {J.-L.}\
  \bibnamefont {Sauvajol}}, \bibinfo {author} {\bibfnamefont {L.}~\bibnamefont
  {Dubois}}, \bibinfo {author} {\bibfnamefont {F.}~\bibnamefont {Lefloch}},\
  and\ \bibinfo {author} {\bibfnamefont {F.}~\bibnamefont {Duclairoir}},\
  }\bibfield  {title} {\enquote {\bibinfo {title} {Growth protocols and
  characterization of epitaxial graphene on {SiC} elaborated in a graphite
  enclosure},}\ }\href
  {https://doi.org/https://doi.org/10.1016/j.physe.2015.07.022} {\bibfield
  {journal} {\bibinfo  {journal} {Physica E: Low-dimensional Systems and
  Nanostructures}\ }\textbf {\bibinfo {volume} {75}},\ \bibinfo {pages} {7--14}
  (\bibinfo {year} {2016})}\BibitemShut {NoStop}%
\bibitem [{\citenamefont {Ohresser}\ \emph {et~al.}(2014)\citenamefont
  {Ohresser}, \citenamefont {Otero}, \citenamefont {Choueikani}, \citenamefont
  {Chen}, \citenamefont {Stanescu}, \citenamefont {Deschamps}, \citenamefont
  {Moreno}, \citenamefont {Polack}, \citenamefont {Lagarde}, \citenamefont
  {Daguerre}, \citenamefont {Marteau1}, \citenamefont {Scheurer}, \citenamefont
  {Joly}, \citenamefont {Kappler}, \citenamefont {Muller}, \citenamefont
  {Bunau},\ and\ \citenamefont {Sainctavit}}]{ohresser2014deimos}%
  \BibitemOpen
  \bibfield  {author} {\bibinfo {author} {\bibfnamefont {P.}~\bibnamefont
  {Ohresser}}, \bibinfo {author} {\bibfnamefont {E.}~\bibnamefont {Otero}},
  \bibinfo {author} {\bibfnamefont {F.}~\bibnamefont {Choueikani}}, \bibinfo
  {author} {\bibfnamefont {K.}~\bibnamefont {Chen}}, \bibinfo {author}
  {\bibfnamefont {S.}~\bibnamefont {Stanescu}}, \bibinfo {author}
  {\bibfnamefont {F.}~\bibnamefont {Deschamps}}, \bibinfo {author}
  {\bibfnamefont {T.}~\bibnamefont {Moreno}}, \bibinfo {author} {\bibfnamefont
  {F.}~\bibnamefont {Polack}}, \bibinfo {author} {\bibfnamefont
  {B.}~\bibnamefont {Lagarde}}, \bibinfo {author} {\bibfnamefont {J.-P.}\
  \bibnamefont {Daguerre}}, \bibinfo {author} {\bibfnamefont {F.}~\bibnamefont
  {Marteau1}}, \bibinfo {author} {\bibfnamefont {F.}~\bibnamefont {Scheurer}},
  \bibinfo {author} {\bibfnamefont {L.}~\bibnamefont {Joly}}, \bibinfo {author}
  {\bibfnamefont {J.-P.}\ \bibnamefont {Kappler}}, \bibinfo {author}
  {\bibfnamefont {B.}~\bibnamefont {Muller}}, \bibinfo {author} {\bibfnamefont
  {O.}~\bibnamefont {Bunau}},\ and\ \bibinfo {author} {\bibfnamefont
  {P.}~\bibnamefont {Sainctavit}},\ }\bibfield  {title} {\enquote {\bibinfo
  {title} {Deimos: A beamline dedicated to dichroism measurements in the
  350--2500 ev energy range},}\ }\href@noop {} {\bibfield  {journal} {\bibinfo
  {journal} {Review of Scientific Instruments}\ }\textbf {\bibinfo {volume}
  {85}},\ \bibinfo {pages} {013106} (\bibinfo {year} {2014})}\BibitemShut
  {NoStop}%
\bibitem [{\citenamefont {Thole}\ \emph {et~al.}(1992)\citenamefont {Thole},
  \citenamefont {Carra}, \citenamefont {Sette},\ and\ \citenamefont {van~der
  Laan}}]{thole_x-ray_1992}%
  \BibitemOpen
  \bibfield  {author} {\bibinfo {author} {\bibfnamefont {B.~T.}\ \bibnamefont
  {Thole}}, \bibinfo {author} {\bibfnamefont {P.}~\bibnamefont {Carra}},
  \bibinfo {author} {\bibfnamefont {F.}~\bibnamefont {Sette}},\ and\ \bibinfo
  {author} {\bibfnamefont {G.}~\bibnamefont {van~der Laan}},\ }\bibfield
  {title} {\enquote {\bibinfo {title} {X-ray circular dichroism as a probe of
  orbital magnetization},}\ }\href
  {https://doi.org/10.1103/PhysRevLett.68.1943} {\bibfield  {journal} {\bibinfo
   {journal} {Physical Review Letters}\ }\textbf {\bibinfo {volume} {68}},\
  \bibinfo {pages} {1943--1946} (\bibinfo {year} {1992})}\BibitemShut {NoStop}%
\end{thebibliography}%

\newpage

\newpage

\newpage

\newpage

\newpage

\end{document}